\documentclass{article}

\usepackage[preprint,nonatbib]{neurips_2020}

\usepackage{echodefs}
\usepackage[title]{appendix}
\usepackage{todonotes}
\usepackage{dashbox}
\usepackage{wrapfig}
\usepackage{multirow}

\newcommand{\calL}{\mathcal{L}}

\newcommand{\calW}{\mathcal{W}}

\newcommand{\down}{\mathsf{D}}
\newcommand{\up}{\mathsf{U}}

\newcommand{\prob}{\mathbb{P}}
\newcommand{\E}{\mathbb{E}}

\newcommand{\vx}{\vec{x}}

\newcommand{\Tnusig}{T_\nu^{\sigma(z)}}

\definecolor{applegreen}{rgb}{0.55, 0.71, 0.0}
\definecolor{armygreen}{rgb}{0.29, 0.33, 0.13}
\definecolor{azure}{rgb}{0.0, 0.5, 1.0}
\definecolor{bittersweet}{rgb}{1.0, 0.44, 0.37}
\definecolor{carrotorange}{rgb}{0.93, 0.57, 0.13}
\definecolor{darkcoral}{rgb}{0.8, 0.36, 0.27}
\definecolor{dandelion}{rgb}{0.94, 0.88, 0.19}

\def\d{\delta} 
 
\def\e{{\epsilon}}

\def\norm#1{\|#1\|}

\newcommand{\ii}{\mathrm{i}}

\newcommand{\pdpd}[2]{\frac{\partial #1}{\partial #2}}

\newcommand{\alphry}{\alpha^{(r)}_{\nu,k}(y)}

\newcommand{\hthetarxi}{\hat\vartheta^{(r)}_{\nu,k}(\xi)}

\newcommand{\mstrut}[1]{\mbox{\rule{0mm}{#1}}}

\renewcommand{\ii}{\mathrm{i}}

\usepackage[utf8]{inputenc} %
\usepackage[T1]{fontenc}    %
\usepackage{url}            %
\usepackage{booktabs}       %
\usepackage{amsfonts}       %
\usepackage{nicefrac}       %
\usepackage{microtype}      %

\usepackage{hyperref}
\hypersetup{
  colorlinks,
  citecolor=blue,
  linkcolor=red,
  urlcolor=blue}

\usepackage[
backend=biber,
style=numeric,
citestyle=numeric,
]{biblatex}

\title{Learning the Geometry of Wave-Based Imaging}

\author{%
  Konik Kothari \\
  UIUC\\
  \texttt{kkothar3@illinois.edu} \\
  \And
  Maarten de Hoop\\
  Rice University\\
  \texttt{mvd2@rice.edu} \\
  \And
  Ivan Dokmani\'{c}\\
  University of Basel\\
  \texttt{ivan.dokmanic@unibas.ch} \\
}

\begin{document}

\maketitle

\begin{abstract}
We propose a general deep learning architecture for wave-based imaging problems. A key difficulty in imaging problems with varying background wave speed is that the medium ``bends'' the waves differently depending on their position and direction. This space-bending geometry makes the equivariance to translations of convolutional networks an undesired inductive bias.
We build an interpretable architecture based on wave physics, as captured by the Fourier integral operators (FIOs). FIOs appear in the description of a wide range of wave-based imaging modalities, from seismology and radar to Doppler and ultrasound. Their geometry is characterized by a canonical relation which governs the propagation of singularities. We learn this geometry via optimal transport in the wave packet representation. The proposed FIONet performs significantly better than the usual baselines on a number of inverse problems, especially in out-of-distribution tests.
\end{abstract}

\begin{wrapfigure}{r}{0.47\textwidth}
    \centering
    \vspace{-4mm}
    \includegraphics[width=.45\textwidth]{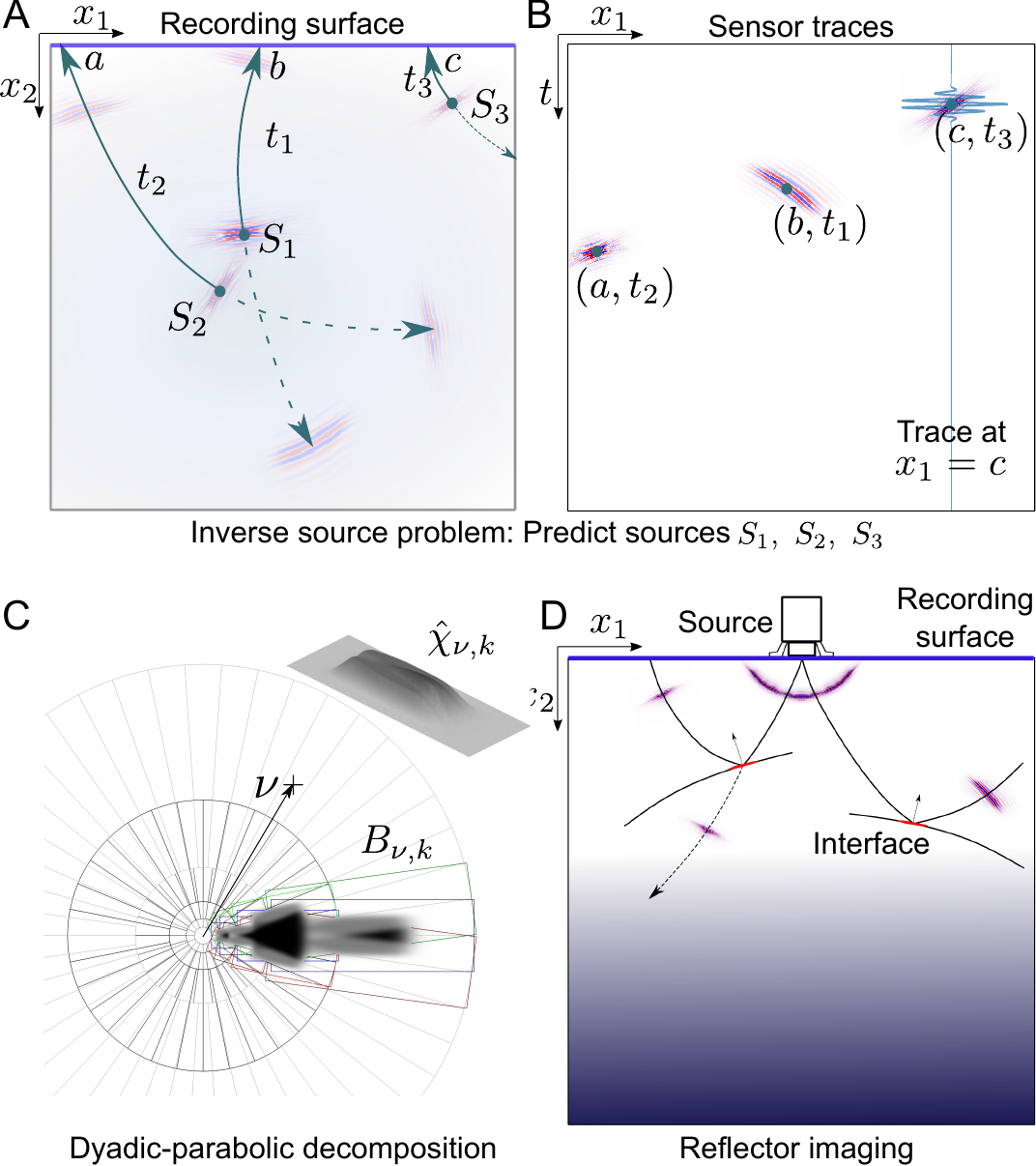}
    \caption{\textit{A:} Wave packets $S_1, S_2$ and $S_3$ are recorded at the surface at times $t_1$, $t_2$ and $t_3$. \textit{B:} The sensor trace is used to image the sources. \textit{C:} 
    Wave packets are localized in frequency by directional bandpass filters $\chi_{\nu, k}^2$. 
    \textit{D:} In reflection imaging, reflections of waves are recorded on the surface (see Appendix \ref{app:add_results}).
    }
    \vspace{-2mm}
    \label{fig:geom_fig}
\end{wrapfigure}

\section{Introduction}

We propose a deep learning approach for wave-based imaging. A simple intuition for imaging with waves can be gleaned from Figure \ref{fig:geom_fig}. Elementary wave packets propagate from where they are created (a source, \ref{fig:geom_fig}A), and then possibly scattered (an interface, \ref{fig:geom_fig}D), to where they are sensed. Where and when a wave packet arrives at a sensor (\ref{fig:geom_fig}B) depends on its orientation and position and the  geometry associated with the background wave speed. To the first approximation, imaging is accomplished by routing the wave packets back where they were created or scattered. Our approach leverages this routing geometry, central to applications from medical thermoacoustic tomography to reflection seismology.

We consider estimating an image $v$ from measurements $u$ obtained by a forward operator $A_\sigma$,
\begin{equation}
  \label{eq:inverse-problem}
  u = A_\sigma v ~ (+ \text{errors}).
\end{equation}
The unknown $v$ could represent the distribution of interfaces (discontinuities) in an otherwise smooth medium. The forward operator (and, hence, its inverse) is parameterized by $\sigma$. Here, $\sigma$ is associated with the parameter of the medium, the background wave speed.
Elementary wave packets travel and bend along curved rays or geodesics defined by the background wave speed, which thus defines a ``geometry''; see Figure \ref{fig:geom_fig}. We do not assume that $\sigma$ is known so we only know $A_\sigma$ up to a class.

We aim to approximate the inverse of \eqref{eq:inverse-problem} by a neural network $f$ trained with a loss $L$. The central question is how to design $f$ and $L$. Our approach is based on the physics of wave propagation as captured by the Fourier integral operators (FIOs) with a loss based on optimal transport.
FIOs describe a diverse set of imaging modalities including reflection seismology~\cite{troot2012linearized,fang2020parsimonious}, thermoacoustic~\cite{ku2005thermoacoustic,stefanov2011thermoacoustic} and photoacoustic~\cite{elbau2012reconstruction} tomography, radar~\cite{ambartsoumian2013class,quinto2011local,cheney2009fundamentals}, and single photon emission computed tomography~\cite{felea2011microlocal}, modeling both forward and inverse maps.

The principal aim of physics-based deep learning is to improve generalization. Common approaches use a (known) forward operator as a component in unrolled iterative reconstruction schemes~\cite{yaman2020self}, or impose it via automatic differentiation~\cite{raissi2017physics,raissi2017physicsII} to ensure data consistency when estimating the inverse from data. Instead of treating a \emph{particular} forward model as a component, we model the internal mechanics of an entire \emph{class} of (inverse) operators related to waves. FIOs and their geometry are the central objects in our design. Nonetheless, the resulting architecture is general and may be used to address a wide range of inverse problems. Since it is matched to physics, it generalizes strongly to examples that are well out of training distribution.

An FIO $F_{\sigma}$ maps the input $u \in L_2(\R^2)$ (for example, a record of pressure time traces) to its output $F_{\sigma}[u]$ (for example, an image of a human brain), as
\begin{equation} \label{eq:fio}
    F_{\sigma}[u](y) = \frac{1}{(2\pi)^2} \int_{\R^2} a(y, \xi) \hat{u}(\xi) e^{iS_{\sigma}(y,\xi)} d\xi,
\end{equation}
where $\hat{u}$ denotes the Fourier transform of $u$, $a(x, \xi)$ is called the symbol of $F_{\sigma}$, and $S_{\sigma}(x,\xi)$ is a suitable phase function (cf. Section \ref{subsec:diffeo}). (Hereafter we suppress $\sigma$ for simplicity.) FIOs are a natural extension of convolutions. If $S(x, \xi) = \langle x, \xi \rangle$ and $a(x, \xi) \equiv \hat{a}(\xi)$, \eqref{eq:fio} is indeed a convolution; it models simple deblurring or denoising. Allowing a general $a(x, \xi)$ makes it a pseudodifferential operator; these appear as approximate solutions of elliptic PDEs~\cite{taylor2012pseudodifferential} or normal operators of imaging~\cite{jin2017deep}. For a general phase $S(x, \xi)$, $F$ becomes powerful: it can deform the domain in an orientation-dependent way. This models approximate solutions of hyperbolic PDEs and therefore wave propagation. The geometry of an FIO (Figure \ref{fig:geom_fig}) is completely captured in its phase $S(x, \xi)$.

\subsection{Our results}

Our architecture design, based on discretization of FIOs~\cite{candes2007fast,de2013multiscale}, enable strong out-of-distribution generalization without any special training schemes or meta-learning paradigms~\cite{balaji2018metareg,li2017deeper,li2018learning,kothari2018random}. This is essential to imaging in exploratory sciences and medical applications where not having out-of-distribution generalization can be disastrous~\cite{antun2020instabilities}.
A key ingredient that allows this is a neural network that learns geometry---the \textit{routing network}. This network is interpretable in that its output corresponds to how wave packets are routed. Another key ingredient to learning this geometry is a training strategy and an optimal transport based loss function. The routing network warps pixel grids and never looks at pixel intensities. Hence, once trained, it is data-independent.

\subsection{Relation to prior work}

Existing physics-based approaches include substituting forward models into unrolled networks and applying auto-differentiation to spatiotemporal fields parameterized by neural networks~\cite{bubba2019learning,raissi2017physics,raissi2017physicsII}. In either case the forward operator should be known in closed from and simple to implement; neither is true in our case. The most popular choice for end-to-end learning  in imaging are convolutional neural networks (CNNs). There is a vast number of papers on supervised learning for inverse imaging problems; we mention a small selection~\cite{shen2017deep,schwab2018real,rivenson2017deep,jin2017deep}. 
CNNs are (approximately) translation-covariant and they excel in problems that are classically solved by filtering. Examples are deblurring, denoising or the inverting Radon transform which becomes a Fourier multiplier upon a composition with its adjoint. Versatile architectures like the U-Net~\cite{ronneberger2015u} can be applied to more general problems, but as we show here this has limitations when the physics is not captured. A related issue is the lack of interpretability: it is not straightforward to associate different parts of a CNN  with corresponding physical processes.

In the context of waves, architecture based on wavelet transforms~\cite{fan2019bcr} were applied to various imaging modalities~\cite{fan2019solving,fan2019OT, fan2019TT}. It is however not clear whether the various components indeed generalize out-of-distribution or how they compare to standard high-quality baselines such as the U-Net, which performs surprisingly well on simple generalization tasks. Finally, we point out the work on meta learning for Calder\'on-Zygmund operators~\cite{feliu2020meta}; our $\sigma$ is similar to their parameterizations.

\section{Imaging with Fourier integral operators} \label{sec:fio}

The geometry of wave packet routing by FIOs becomes apparent upon decomposing the input via a bank of directional bandpass filters. We now discuss a particular choice of directional filters which not only reveals the geometry, but also leads to a computationally efficient representation.

\subsection{Filtering $u$ to a box in the dyadic parabolic tiling of Fourier space} \label{subsec:dyadic}

It has been shown in~\cite{smith1998parametrix} that for wave propagators the so-called dyadic-parabolic tiling~\cite{stein1993harmonic} of the Fourier space  shown in Figure \ref{fig:geom_fig}C is optimal. Such a tiling divides the Fourier space into overlapping boxes, $B_{\nu,k}$, where the length of the box is approximately square of its width (Figure \ref{fig:geom_fig}C). 
The boxes are indexed by $\nu, k$ where $\nu \in \S^1$ denotes the orientation and $k$ the scale. We construct a sequence of smooth directional bandpass filters in the Fourier domain $\hat\chi^2_{\nu,k}$ supported on $B_{\nu,k}$, which form a partition of unity,
\( \label{eq:part_of_unity}
   \hat{\chi}_{0}^2(\xi) + \sum_{k \ge 1} \sum_{\nu}
   \hat{\chi}_{\nu,k}^2(\xi) = 1.
\)
We introduce the directional bandpass components
\(
\hat{u}_{\nu,k}(\xi) = \hat{\chi}^2_{\nu, k}(\xi)\hat{u}(\xi).
\)
Note that $\hat{u}(\xi) = \sum_{\nu,k}\hat{u}_{\nu,k}(\xi)$ by definition of $\hat{\chi}_{\nu,k}$. We let $k_{\text{min}}$ be the coarsest and $k_{\text{max}}$ the finest scale considered in our network design. A detailed sampling analysis relating $k_{\text{max}}$ to discretization can be found in~\cite{duchkov2010discrete}.

\subsection{Geometry of FIOs: diffeomorphisms} \label{subsec:diffeo}

We now show how the phase function of an FIO characterizes the geometry of routing. The phase $S$ is positive homogeneous of degree $1$ in $\xi$. A Taylor expansion of $S(y,\xi)$ in $B_{\nu,k}$ around $\nu$ is then 
\begin{equation} \label{eq:taylor_phase}
   S(y,\xi) = \left\langle\xi,\pdpd{S}{\xi}(y,\nu)\right\rangle
                    + S_2(y, \xi) + \text{higher order terms}.
\end{equation}
The significance of the dyadic parabolic decomposition is that the second-order term $S_2(y, \xi)$ varies only slowly within a box, so $\exp(\ii S_2(y, \xi))$ can be absorbed in the amplitude $a(y, \xi)$. Following this expansion and ignoring the amplitude $a(y, \xi)$ and the $S_2$ term, for a box-filtered $u_{\nu, k}$, we have
\[
    F u_{\nu, k}(y) \approx  \frac{1}{(2\pi)^2}\int_{\R^2}  \hat{u}_{\nu, k}(\xi) e^{i \langle \xi, \parder{S}{\xi}(y, \nu) \rangle} d\xi = u_{\nu, k} \left( \frac{\partial S}{\partial \xi}(y, \nu) \right)
\]
This means that the imaging operator could be roughly approximated via a simple diffeomorphism, $y \to T_{\nu}(y)= \partial_{\xi} S(y,\nu)$, when $u$ is constrained to a box. This transform is what we aim to learn for all $\nu$s via a neural network we call the routing network.

\subsection{Low-rank separated representations} \label{subsec:lr}

To get a more accurate approximation, we incorporate the amplitude $a(y, \xi)$ and the second-order term $S_2(y, \xi)$ via a low-rank separated representation \cite{de2013multiscale},
\[
   a(y, \xi) \exp\left[\ii S_2(y, \xi) \right] \mathbf{1}_{\nu,k}(\xi)
             \ \approx\ \sum_{r=1}^{R_k} \alphry \hthetarxi ,
\]
where $R_k \sim k/\log(k)$. Multiplications by $\hthetarxi$ act as convolutions, while $\alphry$ correspond to simple (diagonal) spatial multipliers. We can now use \eqref{eq:taylor_phase} in \eqref{eq:fio} to write the action of the FIO on $u$ as

\begin{equation}
\begin{aligned}
   (F u)(y) \
   &\approx \
     \sum_{\nu,k} 
    \sum\limits_{r=1}^{R_{\nu,k}}
     {\color{black}
     \alpha_{\nu,k}^{(r)}(y) 
     }\ 
     \sum\limits_{\xi \in \mathbf{1}_{\nu,k}}
     {\color{black}
     e^{\ii \langle T_\nu(y),\xi\rangle} }
     {\color{black}
     \sum\limits_{\nu',k'}
     \hat{\boldsymbol{\vartheta}}_{\nu,k;\nu',k'}^{(r)}(\xi)
     \hat\chi^2_{\nu',k'}(\xi) \, \hat u(\xi)
     } .
\end{aligned}
 \label{equ:Fuy}
\end{equation}
Note that the $B_{\nu,k}$'s overlap and therefore, we generalize results of ~\cite{andersson2012multiscale} to allow for interaction across the boxes via the innermost sum.

We now denote  
\(
    C_{\nu, k}(u)  :=  u_{\nu, k}, \ 
    H_{\nu, k}^{(r)}(u)  := \sum_{\nu',k'} \boldsymbol{\vartheta}_{\nu,k;\nu',k'}^{(r)} \conv u_{\nu', k'}, \  
    A_{\nu, k}^{(r)}(w) := \alpha_{\nu, k}^{(r)} \ w,
\)
so that, noting that the sum over $\xi$ acts as a change of coordinates according to $T_\nu$,
\begin{equation}
    \label{eq:FuY-op}
    (Fu)(y) 
    \approx \left[ \sum_{\nu, k} \sum_{r = 1}^{R_{\nu, k}} A_{\nu, k}^{(r)} \circ  
    (H_{\nu, k}^{(r)} \circ C_{\nu, k} (u)) \circ T_\nu \right](y).
\end{equation}

\section{FIONet: the architecture for wave-based imaging} \label{sec:arch}

We now explain how the proposed FIONet implements and, importantly, generalizes the building blocks defined in \ref{eq:FuY-op}.
The dyadic-parabolic partition of the frequency space corresponds to the map $C_{\nu, k}$. It is a fundamental property of an FIO strongly tied to the structure of wave propagation. We thus implement it using fixed, non-trainable box filters constructed from PyCurvelab~\cite{pycurvelab}.

We then design neural networks $f_{H, \theta_H}$, $f_{T, \theta_T}$ and $f_{A, \theta_A}$ with parameters $\theta_H$, $\theta_T$, $\theta_A$ such that 
\begin{equation} \label{eq:map_fio_to_networks}
\begin{array}{rclcrcl}
    f_{H, \theta_H} &:& \R^{M \times M \times N_{\text{b}}} \to \R^{M \times M \times N_{\text{b}}  R} 
    && [f_{H, \theta_H}(w)]_{\nu, k}^{(r)} &\approx& H_{\nu, k}^{(r)} (w_{\nu, k}) \\
    f_{T, \theta_T} &:& \R^{p} \times \R^{2} \times \R^{2} \to \R^{2}
    && f_{T, \theta_T}(z, y, \nu) &\approx& T_\nu^{\sigma(z)}(y) \\
    f_{A, \theta_A} &:& \R^{M \times M \times N_{\text{b}} R} \to \R^{M \times M \times N_b}
    && [f_{A, \theta_A}(w)]_{\nu, k} &\approx& \sum_{r = 1}^{R} A_{\nu, k}^{(r)} w_{\nu, k}^{(r)}
\end{array}
\end{equation}
We let $f_{H, \theta_H}$ operate on the entire stack of box-filtered inputs, $u_{\nu,k}$ to enable channel interaction. $A_{\nu, k}^{(r)}$ is implemented via simple multiplication layers. Here $N_b$ is the number of boxes in our tiling and $R := R_{k_{\text{max}}}$ is the maximum number of terms in \eqref{eq:FuY-op}. We denote the full set of trainable FIONet network parameters by $\Theta = (\theta_H, \theta_T, \theta_A)$.
The network output is  
\[
    \mathrm{FIONet}_\Theta[u](y) = \sum_{\nu, k} f_{A, \theta_A}(f_{H, \theta_H}^{\nu, k, :}(u)(f_{T, \theta_T}(y, \nu))),
\]
which maps to \eqref{eq:FuY-op} and Figure \ref{fig:fionet_arch}. Note that we allow the geometry network $f_{T, \theta_T}$ to work with multiple medium parameters $\sigma$ (cf. \eqref{eq:inverse-problem}). We assume that $\sigma$ belongs to a set of plausible medium parameters parametrized by $z \in \mathcal{Z}\subseteq \R^p$ and write  $\sigma$ as $\sigma(z)$ and corresponding $T_\nu$ as $T^{\sigma(z)}_\nu$.

\subsection{Geometry module: warped grids and resampling} \label{subsec:arch_geom}

We begin with the routing network $f_{T, \theta_T}$, the central component of the geometry module that routes wave packets via diffeomorphisms introduced in \eqref{eq:taylor_phase}\footnote{We can learn more general transformations than diffeomorphisms; for example, we can handle caustics~\cite{de2013multiscale}.}.
The routing network calculates for each point $y$ where would a $\xi$ (or $\nu$)-oriented wave packet arrive from. This corresponds to $x = \Tnusig(y)$.

Here we focus on a single wavespeed for simplicity and therefore fix $z=z'$. Appendix \ref{app:HJ_flow} shows preliminary results for learning the geometry for multiple backgrounds.

\begin{figure*}
    \centering
    \includegraphics[width=.92\textwidth]{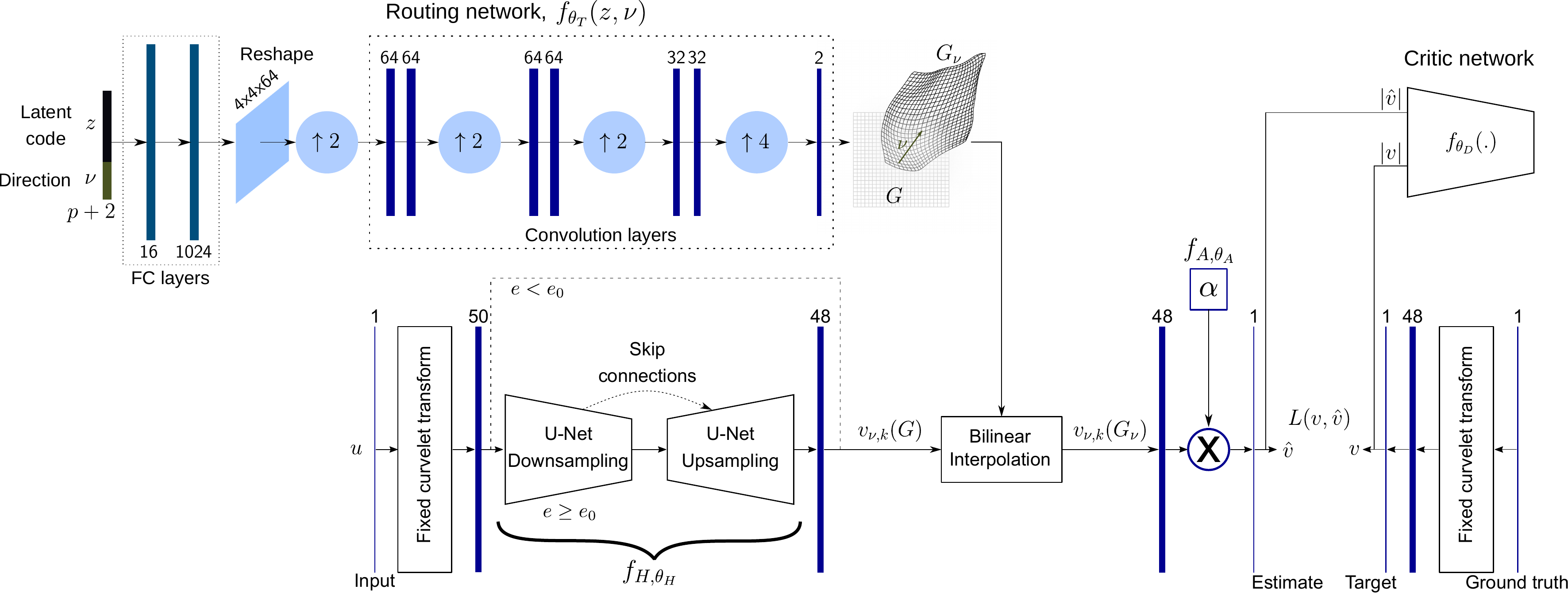}
    \caption{The FIONet. Upsampling blocks use bilinear interpolation.}
    \label{fig:fionet_arch}
\end{figure*}

The input to the geometry module are images defined on the pixel grid 
\(
G := \{ (i/M, j/M) \  : \ (i,j) \in [M]\times[M]\}.
\) 
$G$ represents the pixel centers of an $M\times M$ image $I(G)$. Therefore, even though \eqref{eq:map_fio_to_networks} would suggest to train a fully connected network, $f_T: \R^{p+4} \to \R^2$ note that $y$ is always a point in $G$. Hence, we train a network, $f_T$ that takes in only the latent code $z'$ and $\xi$ and gives the entire warped grid, $G_{\nu} = \{\Tnusig(y) \ :  \ y\in G\}$ in one pass. The architecture of $f_T$ is shown in Figure \ref{fig:fionet_arch}. 

Each channel of the input $I$ to the geometry learning module is associated with a direction $\nu$. We resample each channel, $I_{\nu}$ on the grid $G_\nu$ via bilinear interpolation i.e. the output $I'_{\nu}(G) = I_{\nu}(G_\nu)$, where we zero-fill if points in $G_\nu$ lie outside $G$. Note that the routing network does not ``see'' the images but only warps the grids they are defined over. Hence, once trained it is data-independent. Further note that even though resampling via bilinear interpolation is a linear operation, the morphing of grid $G$ into $G_\nu$ can be significantly nonlinear (see Figure \ref{fig:isp_diff}).

\subsection{Learning diffeomorphisms via optimal transport} \label{subsec:learn_geom}

\begin{figure}
  \centering
    \includegraphics[width=0.75\textwidth]{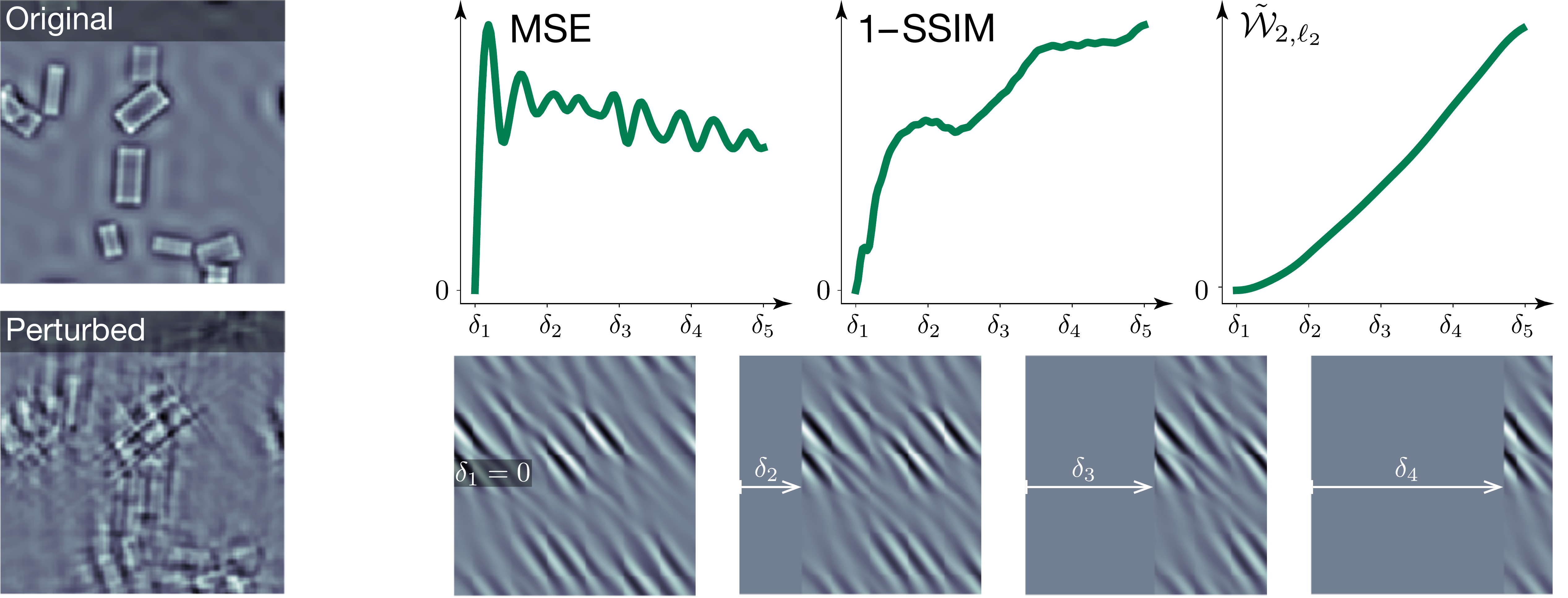}
    \caption{\textit{Left:} Small shifts of the $(\nu, k)$ channels introduce strong distortion. \textit{Right:} Comparison of metrics between oscillatory images: MSE, $\mathrm{SSIM}(|x|, |x_{\delta}|)$ and smooth $\tilde{\calW}_{2,\ell_2}$ between ${|x|}/\norm{x}_1$ and ${|x_{\delta}|}/\norm{x_\delta}_1$ via Sinkhorn iteration ~\cite{sinkhorn1967concerning}.}
    \label{fig:mse}
\end{figure}

Directly training the routing network requires training pairs $\set{(y_i, \xi_i), (x_i)}$. However, we only have a labeled set of input--output images $\set{(u_i, v_i)}$ \eqref{eq:inverse-problem}. The routing information is implicit in those images and should be inferred using a suitable training strategy. The idea of warping grids was also introduced in~\cite{jaderberg2015spatial} but that acts independently on each channel of the features. Here we require that different $T_\nu$s are coordinated so as to get the final reconstruction (see Figure \ref{fig:mse}A).

We train in two stages: first we only train the geometry module which captures the bulk of the physics. This yields a coarse approximation of the imaging operator which routes the wave packets to correct locations. This stage is important to prevent the U-Net from overfitting the training data. We train $f_\theta(u) = \sum_{\nu,k}u_{\nu,k}(f_T(z',\nu))$ such that an appropriate loss metric $\calL_1(v, f_\theta(u))$ is minimized.

Figure \ref{fig:mse} shows that the MSE between translated $u_{\nu, k}$ is not a good metric. Since $u_{\nu, k}$ are oscillatory it has many local minima; this is also known as cycle skipping~\cite{cycleskip1,cycleskip2,cycleskipOT}. SSIM is smoother but varies very little and therefore does not give ``good'' gradients for training. A natural optimal transport metric based on entropically smoothed $\calW_{2,\ell_2}$ gives consistently increasing distance metric.

While smoothed Wasserstein metrics have been used for imaging inverse problems~\cite{adler2017learning} via the iterative Sinkhorn-Knoop algorithm~\cite{cuturi2013sinkhorn}, we find that backpropagating through the iterates is unstable. We thus adopt the method of \cite{dukler2019wasserstein} and weaken the loss to an unsupervised one: instead of matching $u_i$s to $v_i$s in a paired fashion, we match their marginal distributions, $\prob_v$ and $\prob_{f_{\theta_T}(u)}$ while, importantly, still using $\calW_{2,\ell_2}$ between images as the ground metric. To this end, we use a critic network $f_{\theta_D}$ that is employed only during stage-1 ``geometric'' training. The critique network estimates the $\calW_{1,\calW_{2,\ell_2}}(\prob_v, \prob_{f_{\theta_T}(u)})$, giving the final learning objective as
\[
\min_{\theta_T} \max_{\theta_D}  \E_{\hat{v}\sim\prob_{f_{\theta_T}(u)}} f_{\theta_D}(|\hat{v}|) - \E_{v\sim\prob_u}f_{\theta_D}(|v|) + 
\lambda \E_{\tilde{v}\sim\prob_{\mathrm{int}}}(\|\nabla f_{\theta_D}(\tilde{v})\|_{\calW_{2,\ell_2}} - 1 )^2 .
\]
where $\prob_{\mathrm{int}}$ is the density generated via linear interpolations of samples between $\prob_{f_T(u)}$ and $\prob_v$. 

Distribution matching synchronizes the diffeomorphisms to produce sharp images. From Figure \ref{fig:mse}, we see that minor misalignments of the $(\nu, k)$ channels strongly distorts the output. However, since this metric only matches the distributions and not actual data pairs it alone does not give us the required result. We run this scheme for a few epochs and then move to the second stage. In this stage, we train the entire FIONet (including the routing network and the U-Net) using only the standard MSE loss. See Figure \ref{fig:fionet_arch} for the architecture details.

\subsection{Modeling the low-rank separated representation by the U-Net} \label{subsec:arch_lr}

Finally, we implement the map $H_{\nu, k}$. We want to use  standard convolutional layers with small filters. However, it is essential to ensure that we can implement the large filters $\boldsymbol{\vartheta}_{\nu, k; \nu', k'}$. Without pooling, the required number of layers is at least linear in the size of the filter; this can lead to excessive numbers of layers. Thus, we use the U-Net. Pooling is known to yield a large ``receptive field'' with $\log_2(\text{filter size})$ layers although we are not aware of any results that show that the U-Net implements arbitrary filters. We give an argument based on the polyphase decomposition in Appendix \ref{app:U-Net_justify}. 

\paragraph{Approximating FIOs by the FIONet}

While the FIONet architecture is more general than FIOs, it is important to show that as a special case it can approximate exact FIOs. We make the following simplifying assumptions: 1) the routing network is implemented using fully connected layers; 2) the U-Net uses regular downsampling instead of max pooling. The first assumption gives us access to standard approximation theorems; in practice it only makes the forward pass slower. 

\begin{theorem}
    \label{thm:fionet}
    There exists a set of weights $\Theta = (\theta_H, \theta_T, \theta_A)$ such that 
    \begin{equation}
    \norm{F[u] - \mathrm{FIONet}_\Theta[u]} = O(2^{-k_{\text{min}}/2}) \norm{u} .   \end{equation}
\end{theorem}
This parallels \cite[Theorem 2.1]{de2013multiscale}. Here, the presumed sampling density in the ``space'' domain is naturally of order $2 \cdot 2^{k_{\text{max}}}$ though an oversampling factor is required.

\section{Experiments} \label{sec:exp}

\begin{figure}
    \centering
    \includegraphics[width=0.87\textwidth]{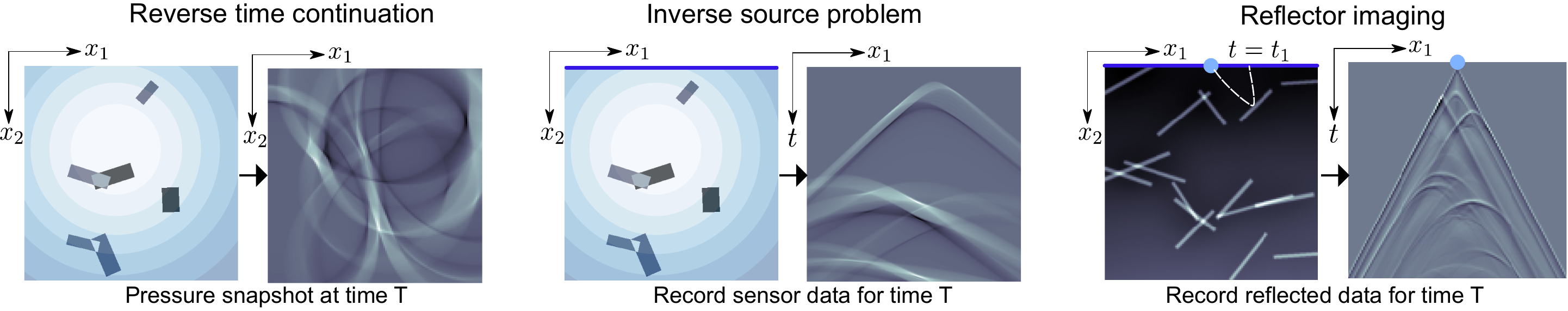}
    \caption{Three inverse problems. \textit{Left:} Reverse time continuation: The initial pressure (boxes) propagates over the shown background wave speed. (b) Inverse source problem: Waves are recorded on the blue sensor line giving sensor traces (c) Reflector imaging: A source (blue dot) sends a pulse that is reflected at the interfaces. The dashed white line show an example ray path.}
    \label{fig:setup}
\end{figure}

We showcase the advantages of learning geometry and the fact that the same network architecture can be applied to various problems. We choose three inverse problems as shown in Figure \ref{fig:setup}: reverse time continuation, inverse source problem, and reflector imaging. We discuss reflector imaging and provide additional results in Appendix \ref{app:add_results}. In all problems, we learn the geometry induced by the background wavespeed directly from data. 
In all our experiments we invoke scale separation. The coarse scale is implicit in the learned geometry. We thus aim to image the fine scales and hence high-pass our target reconstructions.
The dataset and architectures details are given in Appendix \ref{app:training_data_details}. We choose as baseline, the U-Net, arguably the most  successful architecture in imaging ~\cite{schwab2018real,jin2017deep}.
 
\subsection{Reverse time continuation}

\begin{figure}
    \centering
    \includegraphics[width=0.85\textwidth]{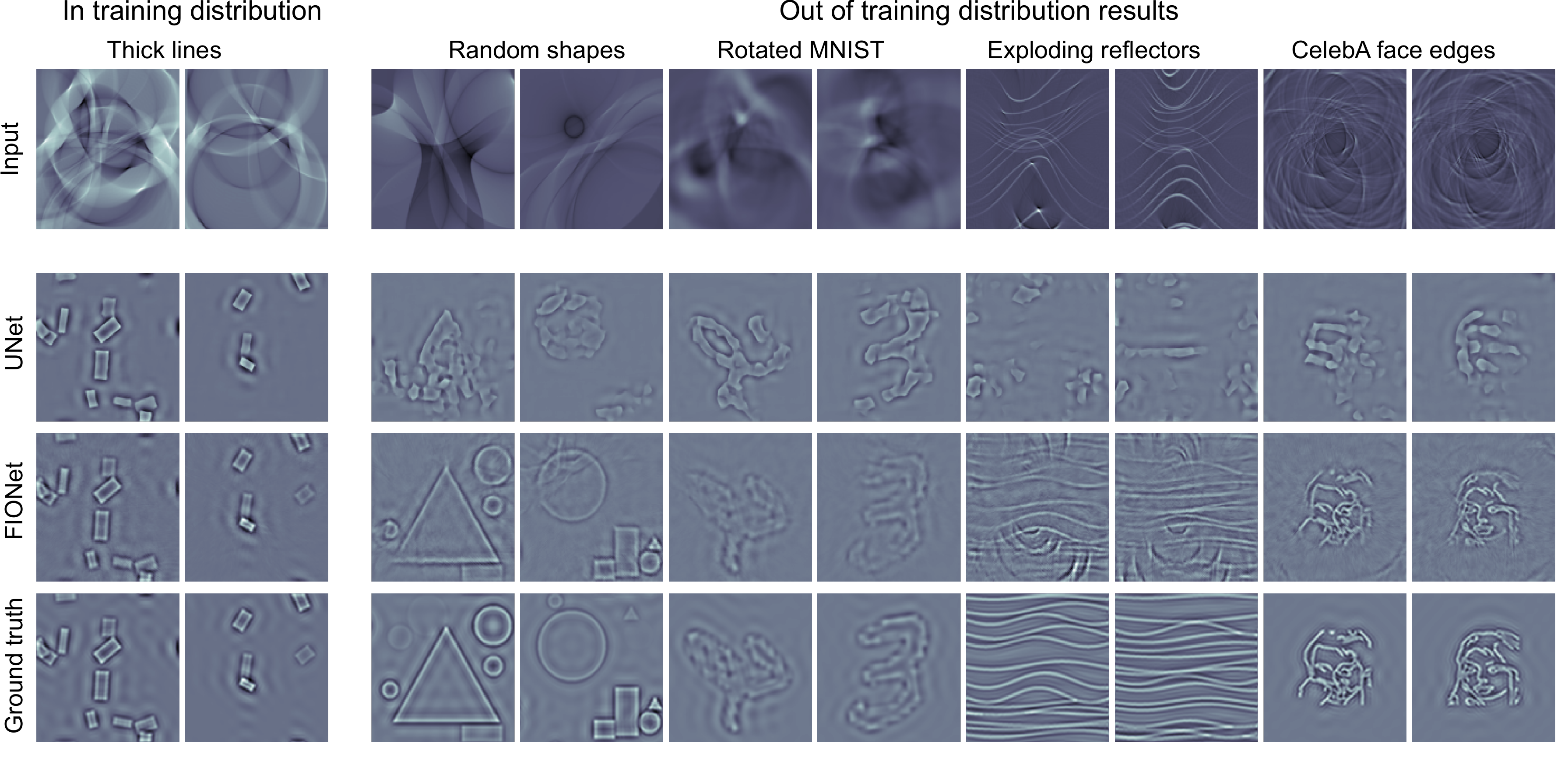}
    \caption{Reverse time continuation results. FIONet performs better in training distribution and is significantly better in our-of-dataset generalization.}
    \label{fig:isp_results}
\end{figure}

\begin{figure}
    \centering
    \includegraphics[width=0.85\textwidth]{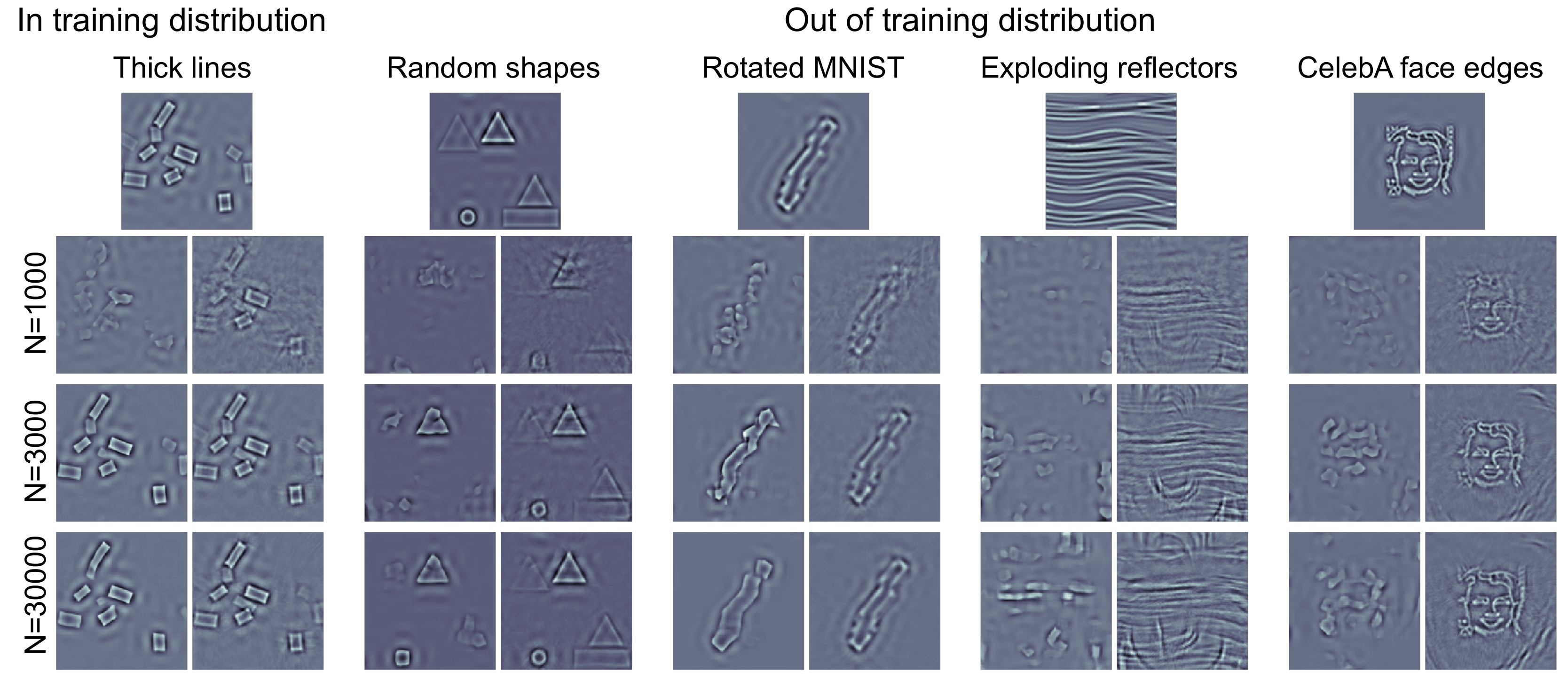}
    \caption{FIONet inductive bias: for each dataset the topmost row shows the ground truth, the left column per dataset shows baseline U-Net results and the right column shows FIONet results. Each row shows the result trained with $N$ samples from the training distribution.}
    \label{fig:ind_bias}
\end{figure}

\begin{figure}
    \centering
    \includegraphics[width=.8\textwidth]{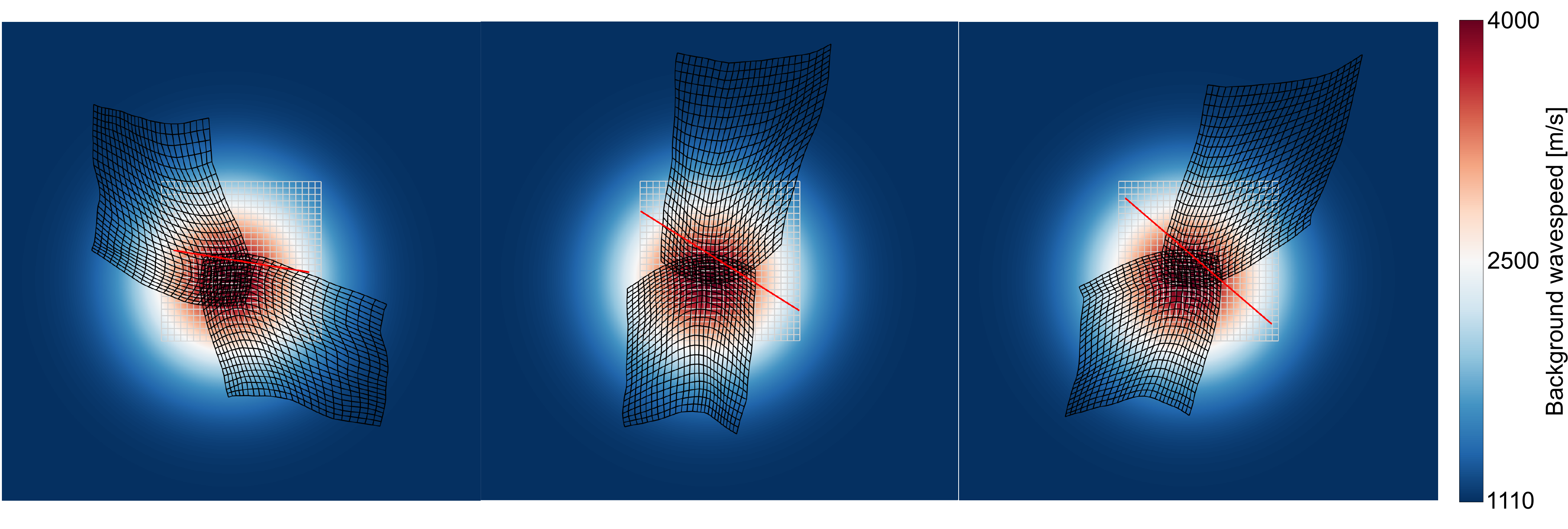}
    \caption{Diffeomorphisms learnt in reverse time continuation by the routing network at different orientations $\nu$ of the wave-packet - red line indicates the orientation of the wave-packet in space. }
    \label{fig:isp_diff}
\end{figure}

In this problem a source pressure field, $p_0$ propagates for time $T$ over an unknown background. We are given the final pressure $p_T$ at $t=T$ and we wish to estimate $p_0$. This problem most intuitively illustrates the geometry of wave propagation (see Figure \ref{fig:isp_diff}). Formally, it corresponds to a sum of two FIOs, one per half-wave propagation (Appendix \ref{app:fio_wave_connect}). We therefore train two copies of $f_{H, \theta_H}$ parallel to model the convolutions (see \eqref{equ:Fuy}), but follow them by a single routing network which outputs 2 warped grids per $\nu$ (see Figure \ref{fig:isp_diff}). The outputs of $f_{H, \theta_{H, 1}}$ and $f_{H, \theta_{H, 2}}$ are resampled on the grids given by the routing network. 
We train on 3000 samples of randomly oriented short thick box sources and test on various distributions. As shown in Figure \ref{fig:isp_results}, in the training distribution the FIONet performs slightly better than the U-Net. In out-of-distribution testing the U-Net seems to synthesize outputs from box-like patterns seen during training and therefore does considerably worse compared to FIONet (see also Table \ref{tab:rtc_summary} in Appendix \ref{app:add_results}). Since we model the transport of wave packets explicitly the convolutional parts of our network can ``focus'' on local enhancement.

\textit{Favorable inductive bias:} Figure \ref{fig:ind_bias} shows that even with a small training set (1000 samples), the FIONet achieves good performance which improves with the dataset size. The U-Net still synthesizes outputs using box-like patterns seen in training set. 

\textit{Interpretability:} Since the routing network explicitly models the geometry of the operator, the $G_\nu$s are physically meaningful estimates (Figure \ref{fig:isp_diff}). The deformed grids clearly show the propagation of the two half-wave solutions (Appendix \ref{app:fio_wave_connect}). We also found that whenever the FIONet did not give reasonable warped grids, the out-of-distribution performance suffered. This suggests that getting the geometry right is indeed central to imaging. This information is not explicitly encoded in any previous architecture. 

\subsection{Inverse source problem} 

In many imaging modalities (photo-acoustic tomography, seismic imaging) sensors are placed at the domain boundary. Instead of having a snapshot of the wavefield at time $T$, we have the time trace at the sensor locations for times $[0,T]$.  The inverse map is modeled by a single FIO~\cite{kuchment2006generalized}.

\begin{figure}
\centering
\includegraphics[width=0.8\textwidth]{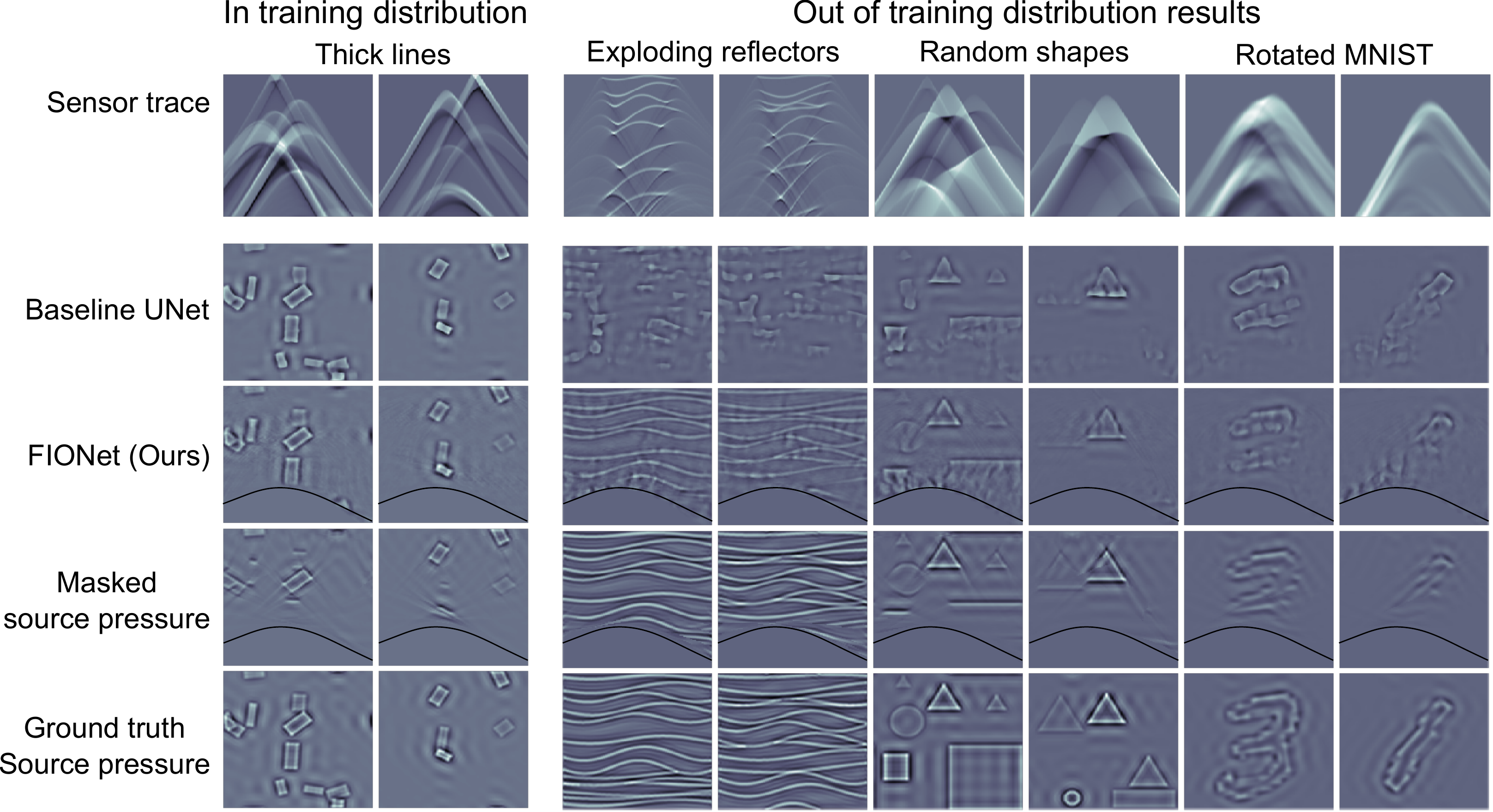}
\caption{Inverse source problem results: We faithfully recover what the sensor sees.}
\label{fig:bs}
\end{figure}

In Figure \ref{fig:bs} we show how the FIONet handles such a scenario. Note that the sensor trace is in the $(x_1, t)$ domain while the source is in the $(x_1, x_2)$ domain. We deliberately choose a background such that not all wave-packets reach the sensor boundary. In Figure \ref{fig:bs}, we only show the interfaces that are ``seen'' by the sensor. We see that these are faithfully recovered by the FIONet. Often in  deep learning approaches to imaging, one claims that since the baseline U-Net reconstructs unseen data as well it is better. However, these networks can be unreliable when tested out-of-distribution~\cite{antun2020instabilities}. Here we aim to be faithful to the physics. 

The FIONet does not predict below the black line which demarcates the ``seen'' and ``unseen'' regions as dictated by the physics. Nonetheless, from the ``seen'' data it still reconstructs more faithfully out-of-distribution than a black-box U-Net even without knowing the background(see Table \ref{tab:isp_summary} in Appendix \ref{app:add_results}).

\section{Conclusion and future work}

We proposed a general architecture, FIONet, and a training strategy for solving inverse problems in wave-based imaging. The routing network---central to our proposal---manifests the geometry of wave propagation and scattering in its output warped grids.
We showed that explicitly learning the geometry enables strong out-of-distribution generalization, outperforming competitive baselines on a variety of imaging problems. This is essential in applications of machine learning in exploratory science.  
FIOs model a remarkable collection of inverse problems, all of which can be addressed with the FIONet. This points to exciting opportunities in applying machine learning to relevant problems in medicine, Earth and planetary sciences, and astronomy.

\clearpage

\begin{ack}
MVdH gratefully acknowledges support from the Department of Energy under grant DE-SC0020345, the Simons Foundation under the MATH + X program, and the corporate members of the Geo-Mathematical Imaging Group at Rice University. ID was supported by the European Research Council Starting Grant 852821---SWING.
\end{ack}

\printbibliography

\begin{appendices}

\section{Additional results} 
\label{app:add_results}

\subsection{Reflector imaging}

\begin{figure}
    \centering
    \includegraphics[width=\textwidth]{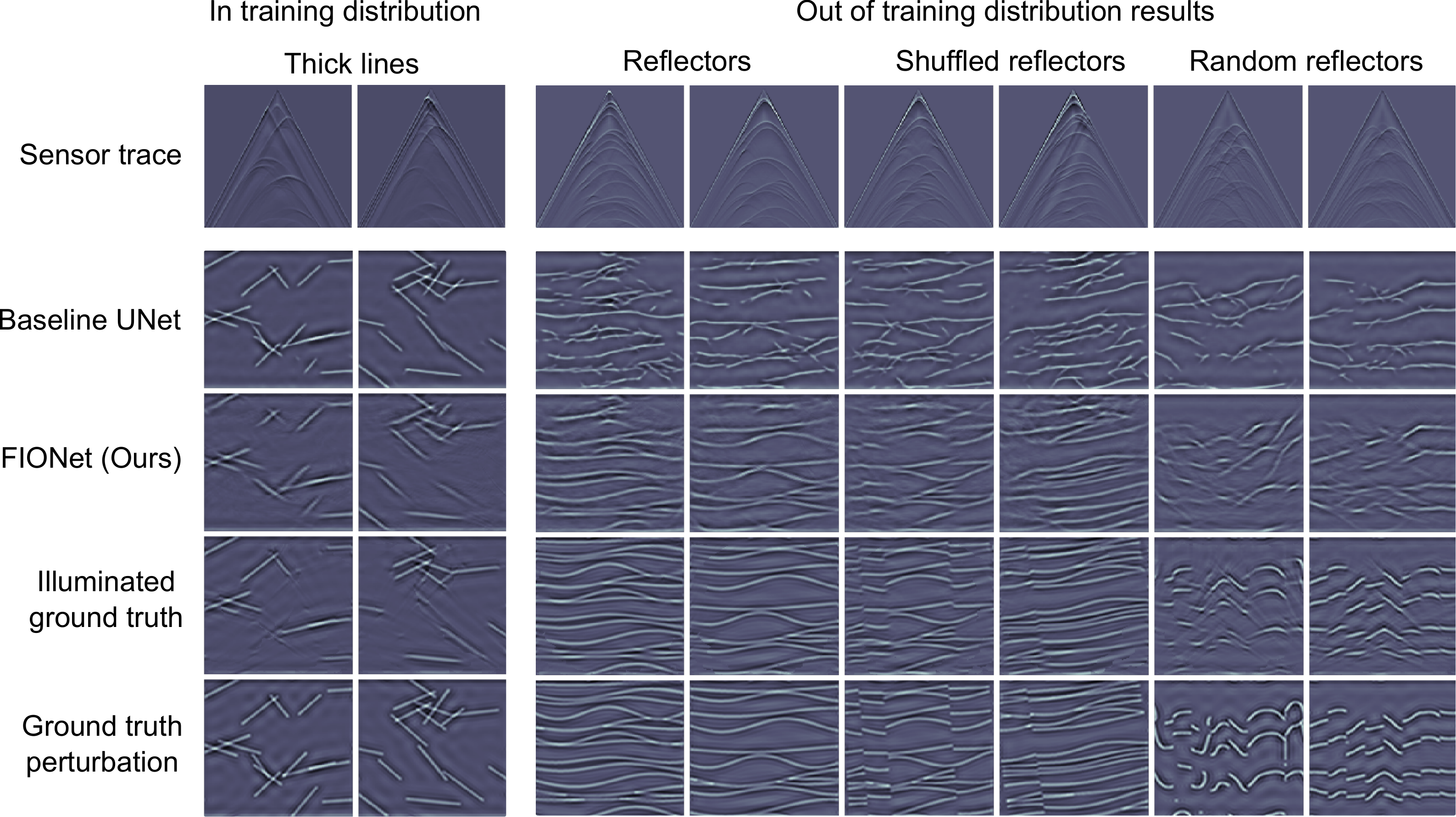}
    \caption{Reflector imaging with a single source, in the second last row, we mask the interfaces that are not illuminated by the source for a fair comparison.}
    \label{fig:full_rtm}
\end{figure}

In reflection seismology, reverse time migration is an important method for imaging the subsurface of a planet. A source pulse is sent from the surface which hits various subsurface interfaces and gets reflected back to the sensors (see Figure \ref{fig:geom_fig} bottom right). 
Again, the map from the sensor trace to the reflector interfaces is an FIO \cite{troot2012linearized} given the background velocity model. We use this to learn the reflector imaging geometry and get reconstructions with a single source \emph{without} knowing the velocity model. We highlight that in practice one uses multiple sources and therefore, with a single source not all interfaces are illuminated. Therefore in Figure \ref{fig:full_rtm}, we also show an ``illuminated-only'' version of the ground-truth as well. This only shows the interfaces that were recorded by the sensors. The results are shown in Figure \ref{fig:full_rtm}. The baseline, same as in other experiments, memorizes data specific patterns and attempts to use them to synthesize the interface distribution. We provide further quantitative results in Table \ref{tab:refl_imaging}. Interestingly, our reconstructions show migration ``smile'' artifacts~\cite{zhu1998smiles} in some out-of-distribution reconstructions which is a well-known phenomenon in seismic imaging when using a single source. 

\subsection{Quantitative results}

We give a quantitative summary of the performance in reverse time continuation, inverse source and reflector imaging problems in Table \ref{tab:rtc_summary}, \ref{tab:isp_summary} and \ref{tab:refl_imaging}. Note that since we apply entropic smoothing to the Wasserstein-$2$, $\tilde{\calW}_{2,\ell_2}$ metric, we see negative values in the metric. The ordering can still be considered maintained meaning that a lower value is better. The second metric we use is the normalized cross-correlation between the images $\chi(x,\hat{x}) = \dfrac{x\cdot\hat{x}}{|x||\hat{x}|}$. For each dataset, we choose a random sample of 20 images and calculate the average performance. Note that for the inverse source and reflector imaging problems, the metrics are calculated based on what data can be plausibly observed by the sensors (see second last row in Figures \ref{fig:bs} and \ref{fig:full_rtm}). For dataset details, refer Appendix \ref{app:training_data_details}.

\begin{table}[tbp] 
\renewcommand{\arraystretch}{1.2} 
\centering
\caption{Reverse time continuation quantitative results. All networks were trained only on the thick lines dataset.}
\begin{tabular}{@{}llrrrrrr}
\toprule
\multicolumn{1}{@{}l}{\multirow{2}{*}{{Dataset}}} &
  \multicolumn{1}{l}{\multirow{2}{*}{{Model}}} &
  \multicolumn{2}{c}{{1000 samples}} & 
  \multicolumn{2}{c}{{3000 samples}} & 
  \multicolumn{2}{r}{{30000 samples}} \\
  \cmidrule{3-4} \cmidrule{5-6} \cmidrule{7-8}
\multicolumn{1}{c}{} &
  \multicolumn{1}{c}{} &
  \multicolumn{1}{c}{$\tilde{\calW}_{2,\ell_2}\downarrow$} &
  \multicolumn{1}{c}{$\chi\ \uparrow$} &
  \multicolumn{1}{c}{$\tilde{\calW}_{2,\ell_2}\downarrow$} &
  \multicolumn{1}{c}{$\chi\ \uparrow$} &
  \multicolumn{1}{c}{$\tilde{\calW}_{2,\ell_2}\downarrow$} &
  \multicolumn{1}{c}{$\chi\ \uparrow$} \\ \midrule
\multirow{2}{*}{\textit{Thick lines}}  & U-Net   & 33.45 & 0.08 & 1.15  & 0.92 & 3.55  & 0.81 \\  \vspace{1mm}
                        & FIONet & 18.47 & 0.72 & -7.40 & 0.97 & -4.89 & 0.94 \\
\multirow{2}{*}{\textit{Shapes}} & U-Net   & 56.84  & \multicolumn{1}{c}{0.02} & 50.80 & 0.21 & 13.71 & 0.53 \\  \vspace{1mm}
                        & FIONet & 9.67   & 0.61 & 1.71  & 0.81 & -4.74 & 0.90 \\ 
\multirow{2}{*}{\textit{Reflectors}} & U-Net   & 28.44  & 0.00                     & 27.94 & 0.07 & 37.09 & 0.08 \\  \vspace{1mm}
                        & FIONet & 11.73  & 0.53                     & 3.52  & 0.63 & 4.01  & 0.56 \\ 
\multirow{2}{*}{\textit{MNIST}}  & U-Net   & 28.44  & 0.04                     & 15.83 & 0.35 & 5.24  & 0.58 \\  \vspace{1mm}
                        & FIONet & 11.73  & 0.69                     & -1.34 & 0.87 & -3.75 & 0.92 \\ 
\multirow{2}{*}{\textit{CelebA faces}}  & U-Net   & 128.30 & 0.02                     & 45.20 & 0.27 & 54.48 & 0.18 \\
                        & FIONet & 68.53  & 0.69                     & 5.93  & 0.83 & 5.63  & 0.85 \\ 
                        \bottomrule
\end{tabular}
\label{tab:rtc_summary}
\end{table}

\begin{table}
\renewcommand{\arraystretch}{1.2} 
\caption{Inverse source problem quantitative results. All networks were trained only on the thick lines dataset.}
\centering
\begin{tabular}{@{}llrr}
\toprule
\multicolumn{1}{@{}l}{Dataset} &
  \multicolumn{1}{l}{Model} &
  \multicolumn{1}{c}{$\tilde{\calW}_{2,\ell_2}\downarrow$} &
  \textbf{$\chi\ \uparrow$} \\ \midrule
\multirow{2}{*}{\textit{Thick lines}}      & U-Net   & 83.28 & 0.60                     \\\vspace{1mm}
                            & FIONet & 17.76 & 0.68                     \\ 
\multirow{2}{*}{\textit{Reflectors}} & U-Net   & 29.06 & 0.38                     \\\vspace{1mm}
                            & FIONet & -4.02 & 0.87                     \\ 
\multirow{2}{*}{\textit{Random shapes}} & U-Net                               & 74.10 & 0.53 \\\vspace{1mm}
                            & FIONet & 33.84 & 0.67 \\ 
\multirow{2}{*}{\textit{MNIST}}      & U-Net   & 63.07 & 0.43 \\
\vspace{1mm}
                            & FIONet & 41.15 & 0.58 \\ \bottomrule
\end{tabular}
\label{tab:isp_summary}
\end{table}

\begin{table}
\caption{Reflector imaging quantitative results. All networks were trained only on the lines dataset.}
\renewcommand{\arraystretch}{1.2} 
\centering
\begin{tabular}{@{}llrr}
\toprule
\multicolumn{1}{@{}l}{Dataset} &
  \multicolumn{1}{l}{Model} &
  \multicolumn{1}{c}{$\tilde{\calW}_{2,\ell_2}\downarrow$} &
  \textbf{$\chi\ \uparrow$} \\ \midrule
\multirow{2}{*}{\textit{Lines}}      & U-Net   & 5.69  & 0.77                     \\\vspace{1mm}
                            & FIONet & 6.74  & 0.75                     \\ 
\multirow{2}{*}{\textit{Reflectors}} & U-Net   & 10.22 & 0.42                     \\\vspace{1mm}
                            & FIONet & 3.17  & 0.74                     \\ 
\multirow{2}{*}{\textit{Shuffled Reflectors}} & U-Net & 8.50 & 0.41 \\\vspace{1mm}
                            & FIONet & 3.09  & 0.72 \\ 
\multirow{2}{*}{\textit{Random Reflectors}} & U-Net & 5.41 & 0.38 \\\vspace{1mm}
                            & FIONet & 4.60   & 0.55 \\ \bottomrule
\end{tabular}

\label{tab:refl_imaging}
\end{table}

\subsection{Stability under noise}
We evaluate the stability of our trained networks under additive noise by testing on 10dB noisy inputs. For a comparison of performance, see Figure \ref{fig:isp_results}. Note that all networks were trained on clean data and then tested on 10dB noisy inputs.

\begin{figure}
    \centering
    \includegraphics[width=\textwidth]{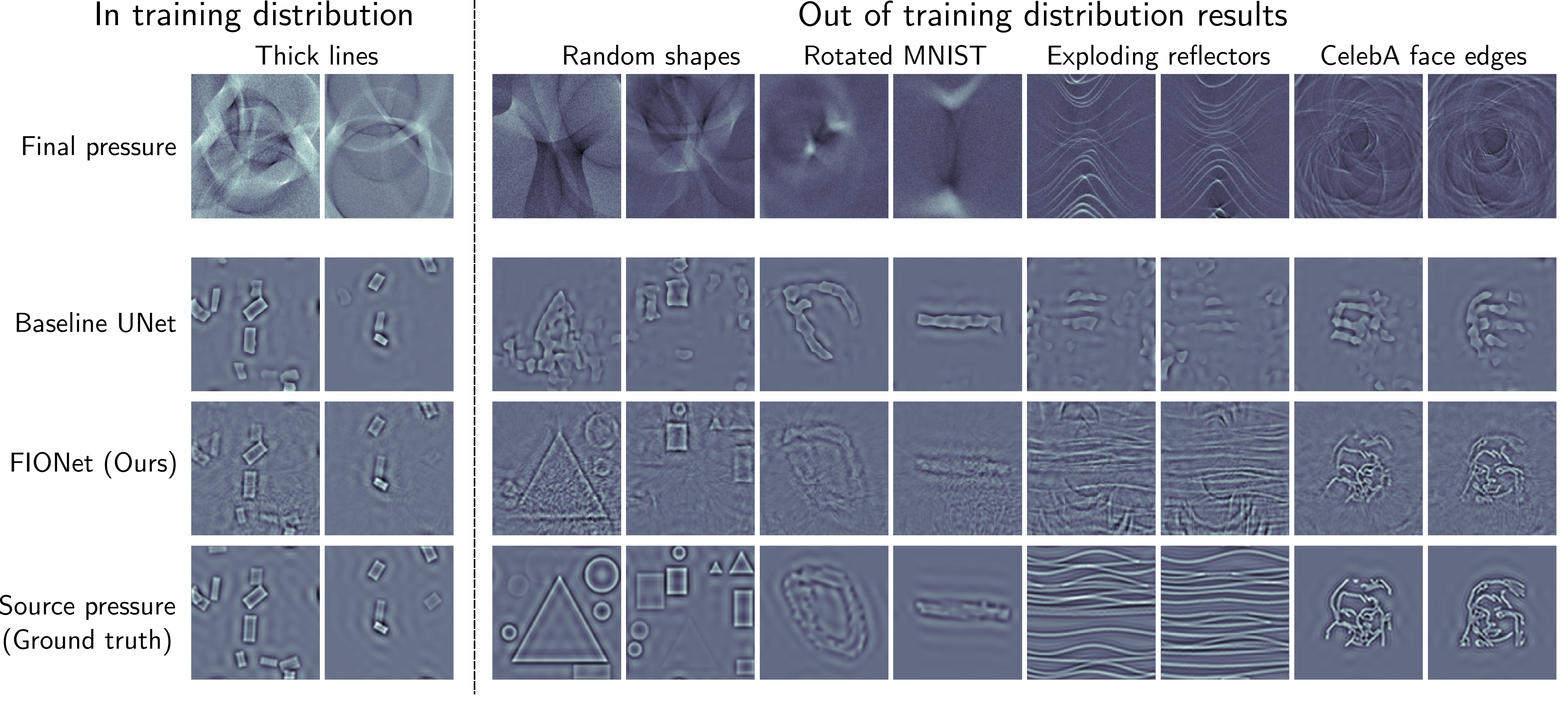}
    \caption{Testing 10dB noisy inputs on networks trained on clean data.}
    \label{fig:isp_noisy}
\end{figure}

\subsection{Routing network output} \label{app:routing_output}

\begin{figure}
\begin{minipage}[l]{0.45\textwidth}
    \centering
    \includegraphics[width=0.95\textwidth]{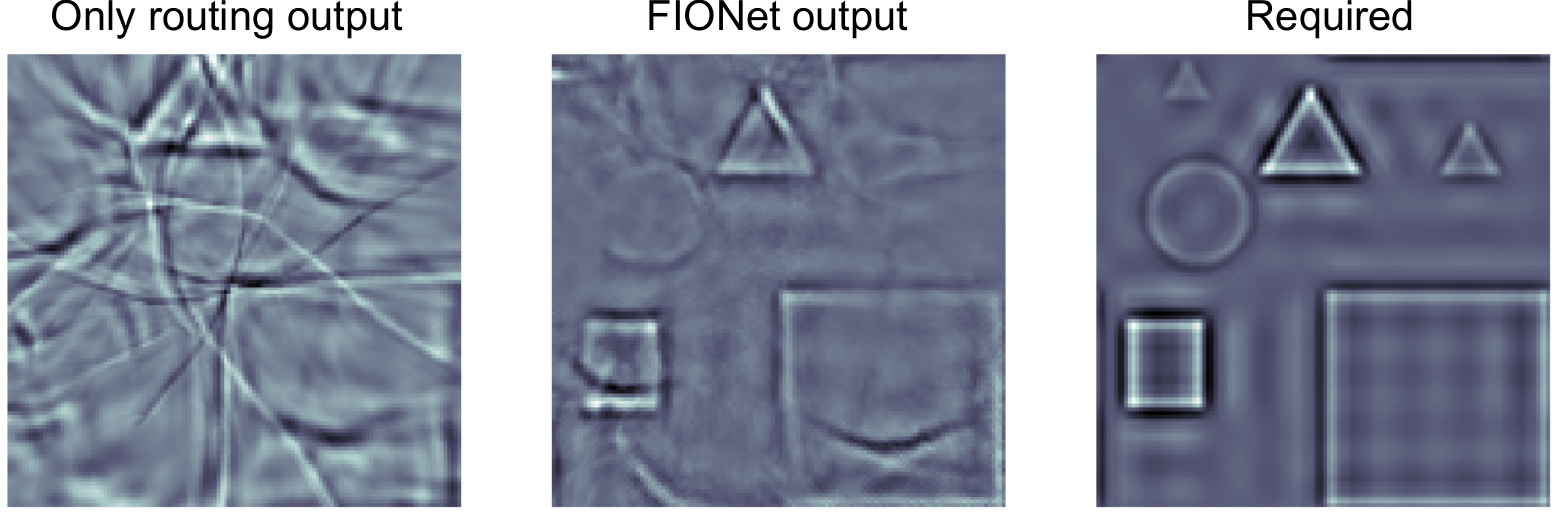}
    \caption{Reverse time continuation: output of the routing network already brings the interfaces close to where they should be.}
    \label{fig:routing}
\end{minipage}
\hspace{2mm}
\begin{minipage}[r]{0.45\textwidth}
    \centering
    \includegraphics[width=0.95\textwidth]{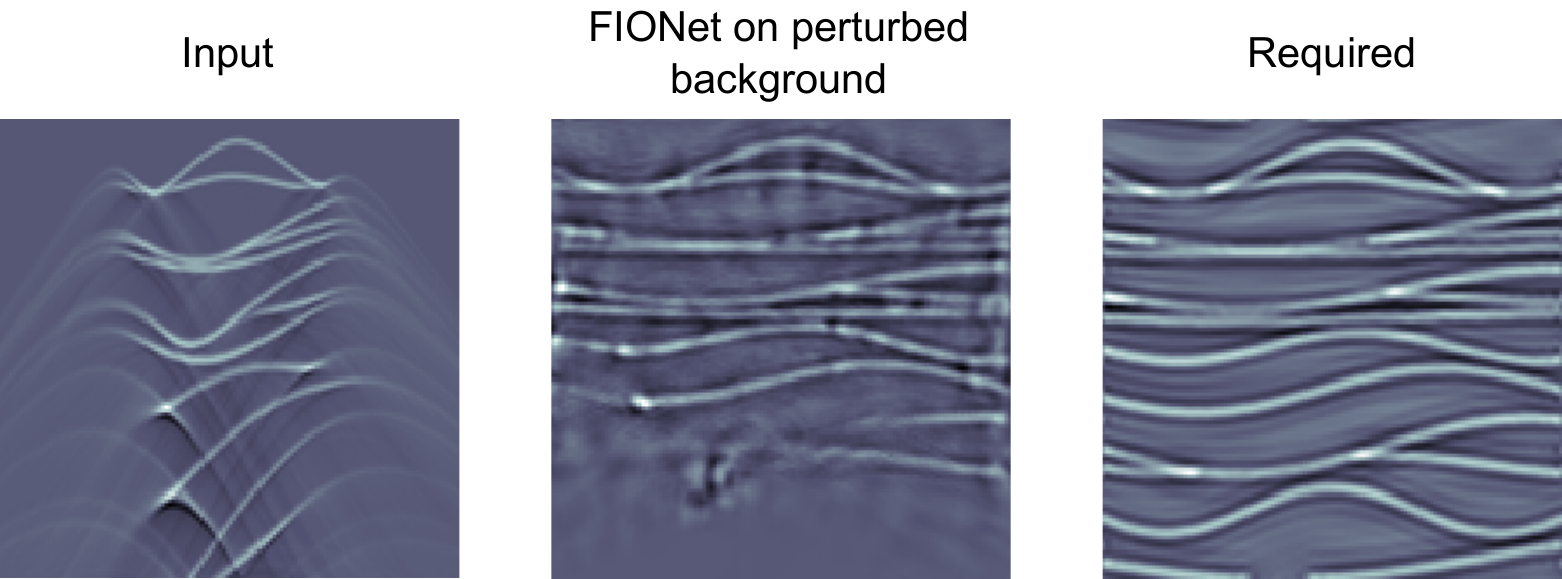}
    \caption{The background wavespeed is perturbed by $5$\% from the training condition and the sensor recorded. On this trace, we see that the reconstruction from the FIONet is stable in that we get the required interfaces.}
    \label{fig:pert_bg}
\end{minipage}
\end{figure}

In order to understand how the routing network warps the image (per $(\nu, k)$ channel), we show an output from the first phase of training. Here the $u_{\nu,k}$s are directly warped over the grids given by the routing network. In Figure \ref{fig:routing} we show the output on the reverse time continuation problem. Note that here two separate half-wave solutions need to synchronize to give the final image (see Appendix \ref{app:fio_wave_connect}). We can see that the routing network already places the interfaces at almost the right locations (obviously with artifacts). The U-Net further filters and enhances the $u_{\nu,k}$s such that after resampling on the  grids given by routing network ultimately gets us close to the required image. Note that the network has never seen any image from this dataset during training. 

\subsection{Robustness against change of background}

In the inverse source imaging problem, we consider a case where the background wave speed at test time has about a $5$\% deviation with respect to the training background wave speed. This is motivated by seismic applications where a $3-5$\% variation in the Earth's mantle wave speed is expected~\cite{lee1984extremal}. We can see in Figure \ref{fig:pert_bg} that recovery is still stable.

\subsection{Figure 1}
\begin{figure}
    \centering
    \includegraphics[width=.8\textwidth]{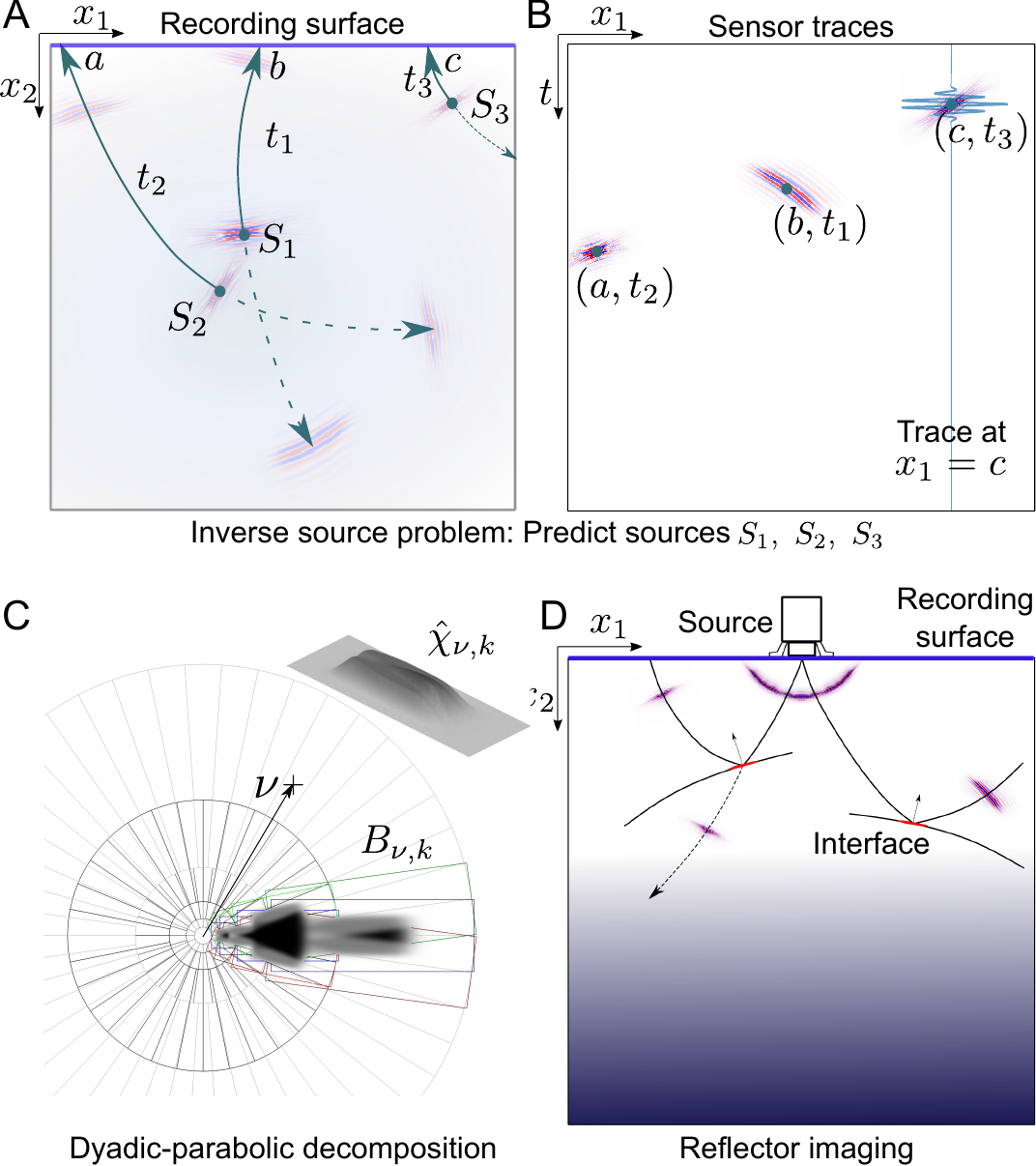}
    \caption{\textit{A:} Wave packets $S_1, S_2$ and $S_3$ are recorded at the surface at times $t_1$, $t_2$ and $t_3$ at the recording surface. Note that the sources travel in two directions(dashed and solid arrows)---two half-wave solutions (see Appendix \ref{app:fio_wave_connect}). In our setup only one half-wave gets recorded.at the recording surface.
    \textit{B:} The sensor trace on the right shows these recordings in the $(x_1,t)$ domain. We see 3 wave-packets at $(a,t_2),\ (b,t_1),\ (c, t_3)$ corresponding to the arrival of $S_2, S_1, S_3$ respectively. We also show a single sensor trace line(blue) overlayed at $x=c$. The orientation and timing of the wave packets in the trace is tied to the orientation and location of the wave packets at $t=0$.
    \textit{C:} Dyadic-parabolic decomposition of phase space. Wave packets are localized in frequency by directional bandpass filters $\hat{\chi}_{\nu, k}^2$ shown in top right. The boxes shown in green,red and blue correspond to $B_{\nu,k}$.
    \textit{D:} In reflection imaging, a source emits a bandlimited pulse that is reflected at interfaces and recorded on the surface (see Appendix \ref{app:add_results}). Notice that the ray bends \textit{and} scatters but we only have the scattered signals. 
    }
    \label{fig:geom_fig_ext}
\end{figure}
\clearpage

\section{FIONet approximates FIOs: Sketch of Proof of Theorem \ref{thm:fionet}} \label{app:U-Net_justify}

The proof of Theorem \ref{thm:fionet} can be decomposed into three parts, namely showing that
\begin{enumerate}[(i)]
    \item $f_{T, \theta_T}$ approximates $T_\nu$;
    \item $f_{A, \theta_A}$ approximates $A_{\nu, k}^{(r)}$ for all $\nu, k$ and $r$;
    \item $f_{H, \theta_H}$  approximates $H_{\nu, k}^{(r)}$ for all $\nu, k$ and $r$.
\end{enumerate}

\paragraph{Part (i)} This follows from the results on universal approximation by deep neural networks~\cite{yarotsky2017error} on noting that $(y, \nu) \mapsto T_\nu(y)$ is smooth in both $y$ and $\nu$. We can thus conclude that for any $\epsilon > 0$ there exists a $\theta_T(\epsilon)$ such that
\begin{equation}
    \label{eq:approx-T}
    \sup_{\nu \in [0, 2\pi), y \in \mathcal{D}}\norm{f_{T, \theta_T(\epsilon)}(y, \nu) - T_{\nu}(y)} \leq \epsilon,
\end{equation}
\noindent where $\mathcal{D}$ is the (compact) computational domain of interest. Since we measure the reconstruction error in $L^2$, this fact alone does not immediately give us the desired estimate. We lean on results from \cite{de2013multiscale} which assume that the diffeomorphisms are implemented perfectly; \eqref{eq:approx-T} does not yield a satisfactory bound on the $L^2$ norm since
\[
    \lim_{\epsilon \to 0} ~ \norm{u(f_{T, \theta_T(\epsilon)}(\, \cdot\, )) - u(T_\nu(\, \cdot \,))}_{L^2} \not\to 0 
\]
if $u$ is allowed to contain arbitrarily high frequencies. Conveniently, this is not true in our case: as we work with discrete pixels, we assume $u$ is adequately bandlimited before sampling. This implies that $u$ is Lipschitz continuous (pending a few technicalities: we assume $u$ is obtained as an inverse Fourier transform of a bandlimited spectrum in $L^1 \cap L^2$) ,
\[
    | u(x_1) - u(x_2) | \leq L_u \norm{x_1 - x_2},
\]
where $L_u$ can be uniformly bounded by $L$ depending on the maximum norm and bandwidth. Further, we only compute the error over a compact domain. We can then write  
\begin{align*}
    \norm{u(f_{T, \theta_T(\epsilon)}(y)) - u(T_\nu(y))}_{L^2(\cal D)}
    &= \norm{u(T_\nu(y) + R(y)) - u(T_\nu(y))}_{L^2({\cal D})} \\
    &\leq L \norm{ R(y) }_{L^2({\cal D})} \\
    &\leq L \sqrt{|\cal D|} \cdot \epsilon
\end{align*}
where the last quantity can indeed be made arbitrarily small since $\cal D$ is fixed.

\paragraph{Part (ii)} This follows trivially since $A_{\nu, k}^{(r)}$ is a simple linear pointwise  multiplication.

\paragraph{Part (iii)} The technical difficulty in part (iii) is proving that a U-Net using small filters can approximate convolutions with $\vartheta_{\nu, k}^{(r)}$. We use a technique from signal processing called the polyphase decomposition~\cite{do2011multidimensional,vetterli2014foundations}.

Consider a single discrete filter kernel $h[\vn]$, $\vn \in \Z^2$ with possibly large but finite support contained within the image. We introduce the 2D $z$-transform as
\begin{equation}
    X(\vz) = \sum_{\vn \in \Z^{2}} x[\vn] \vz^{-\vn},
\end{equation}
with $\vz^{-\vn} := z_1^{-n_1} z_2^{-n_2}$. We split the image $x[\vn]$ into its polyphase components $x_0[\vn] = x[2\vn]$, $x_1[\vn] = x[\vl_1 + 2\vn]$, $x_2[\vn] = x[\vl_2 + 2\vn]$, $x_3[\vn] = x[\vl_3 + 2\vn]$, with $\vl_1 = [1, 0]^T$, $\vl_2 = [0, 1]^T$, $\vl_3 = [1, 1]^T$. Note that $x$ can be assembled from its polyphase components via upsampling and interleaving.

We write
\[
   X_l(\vz) = \sum_{\vn \in \Z^{2}} x_l[\vn] \vz^{-\vn},\quad
   l = 0,1,2,3,
\]
and find that
\[
    X(\vz) = X(z_1, z_2) = X_0(z_1^2, z_2^2) + z_1^{-1} X_1(z_1^2, z_2^2) + z_2^{-1} X_2(z_1^2, z_2^2) + z_1^{-1} z_2^{-1} X_3(z_1^2, z_2^2).
\]
and similarly
\[
    H(\vz) = H(z_1, z_2) = H_0(z_1^2, z_2^2) + z_1^{-1} H_1(z_1^2, z_2^2) + z_2^{-1} H_2(z_1^2, z_2^2) + z_1^{-1} z_2^{-1} H_3(z_1^2, z_2^2).
\]
We aim to compute $y[\vn] = (x \conv h)[\vn]$ or, in the $z$-domain, $Y(\vz) = H(\vz) X(\vz)$. It is possible to write
$Y(\vz)$ as
\begin{equation}
    \label{eq:2d-polyphase-unknown}
    Y(\vz) = 
    \begin{bmatrix}
        1 & z_1^{-1} & z_2^{-1} & z_1^{-1} z_2^{-1}
    \end{bmatrix}
    \begin{bmatrix}
         &  &  &  & \\
         &  &  &  & \\
         &  & \mG(\vz) &  & \\
         &  &  &  &  \\
         &  &  &  & 
    \end{bmatrix}
    \begin{bmatrix}
        X_0(\vz^2) \\
        X_1(\vz^2) \\
        X_2(\vz^2) \\
        X_3(\vz^2)
    \end{bmatrix}.
\end{equation}
such that the a priori non-unique $\mG(\vz)$ depends only on even powers of $z_1, z_2$. That is, the corresponding filters live strictly on the subgrid $\calM = \set{\mM \vn \ : \ \vn \in \Z^2}$, with $\mM = \diag(2, 2)$. When such a filter matrix is followed by regular downsampling, we can exchange the order of downsampling and filtering, by replacing $\vz^2$ by $\vz$.

Some deliberation shows that
\[
    \mG(\vz) =\mH(\vz^2) = 
    \begin{bmatrix}
        H_0(\vz^2) & z_1^{-2} H_1(\vz^2) & z_2^{-2} H_2(\vz^2) & z_1^{-2} z_2^{-2} H_3(\vz^2) \\
        H_1(\vz^2) & H_0(\vz^2) & z_2^{-2} H_3(\vz^2) & z_2^{-2} H_2(\vz^2) \\
        H_2(\vz^2) & z_1^{-2} H_3(\vz^2) & H_0(\vz^2) & z_1^{-2} H_1(\vz^2) \\
        H_3(\vz^2) &  H_2(\vz^2) & H_1(\vz^2) & H_0(\vz^2)
    \end{bmatrix}.
\]

Let
\[
    \vd(\vz) =
    \begin{bmatrix}
        1 & z_1^{-1} & z_2^{-1} & z_1^{-1} z_2^{-1}
    \end{bmatrix}^T \quad \text{and} \quad
    \vx(\vz^2) = 
    \begin{bmatrix}
        X_0(\vz^2) \\
        X_1(\vz^2) \\
        X_2(\vz^2) \\
        X_3(\vz^2)
    \end{bmatrix},
\]
and define the regular downsampling and upsampling operators as
\[
    (\down_2 \vx) [\vn] = \vx[\mM \vn] \quad \text{and} \quad
    (\up_2 \vx) [\vn] =
    \begin{cases}
    \vx[\tfrac{1}{2} \vn] & \vn \in 2 \Z^2 \\
    0 & \text{otherwise}.
    \end{cases}
\]
Noting that $\vx(\vz^2)$ coincides with $\vd(\vz^2) X(\vz^2)$ on $\calM$, we can write
\[
    Y(\vz) = \vd(\vz)^T \bigg(\up_2 \big(\mH(\vz) \down_2 \left( \vd(\vz) X(\vz) \right) \big) \bigg),
\]
with a slight abuse of the $\down_2, \up_2$ notation. This exactly corresponds to a U-Net (with identity activations and no bias) with one downsampling and one upsampling and four channels in between. The filters in the first layer are given as $\vd(z)$ and they are of length at most 2; the filters in the second layer (after the downsampling) correspond to the (shifted) polyphase components of $H(\vz)$ so they are of length about $K/2$ for a filter $h[\vn]$ with support size $K \times K$. Recursively continuing this procedure increases the number of channels by a factor of 4 and halves the filter lengths. We need about $\log_2 K$ downsampling and $\log_2 K$ upsampling layers to implement $h[\vn]$ using filters of size at most $3 \times 3$. This implies that the number of channels in the innermost layer is about $4^{\log_2 K} = K^2$. We note that with ReLU activations a filtering can be standardly written as
\begin{equation}
   h \conv x
   = \left[\begin{array}{cc} I & -I \end{array}\right]
   \left[\begin{array}{r}
           \operatorname{ReLU}(+ \ h \conv x \, ) \\
           \operatorname{ReLU}(- \ h \conv x \, )
        \end{array}
   \right],
\end{equation}
yielding the way to insert ReLU activations in each layer. Thus, on the discretized level, the U-Net architecture can exactly represent the convolutions with $\boldsymbol{\vartheta}^{(r)}_{\nu,k;\nu',k'}$.

With Parts (ii) and (iii), the FIONet reproduces (3.11) in~\cite{andersson2012multiscale} upon eliminating cross channel interaction in the filters of the U-Net. Part (i) provides an estimate of misalignment separately from this. The work of \cite{duchkov2010discrete} provides an estimate for the approximation of \eqref{equ:Fuy} by (3.11) referred to above using numerical analysis which is controlled by an oversampling factor. We absorb the estimate of misalignment in this estimate. We then apply Theorem 4.1 in~\cite{de2009seismic} to obtain the result using curvelets from a tight frame.

\section{FIOs and the wave equation} \label{app:fio_wave_connect}

The Cauchy initial value problem for the scalar wave equation is given
by
\begin{eqnarray} \label{eq:decoupled_equation}
& &P(x,D_x,D_t) u = 0 ,\quad
   P(x,D_x,D_t) = \partial_t^2 + c(x) \left(\mstrut{0.65cm}\right.
   \sum_{j=1}^2 D_{x^j}^2 \left.\mstrut{0.65cm}\right ) \, c(x)
\\
& &u |_{t=0} = h ,\quad \partial_t u |_{t=0} = h' ,
\label{eq:initcond}
\end{eqnarray}
where $D_{x} = -i \pdpd{}{x} \leftrightarrow \xi$ (via the Fourier transform). We summarize how to solve
(\ref{eq:decoupled_equation})-(\ref{eq:initcond}) with the plane-wave
initial value,
\[
   h(x) \equiv 0 ,\quad\ \ h'(x) = \exp[i \langle \xi,x \rangle] ,
\]
where $\xi \in \mathbb{R}^2 \setminus \{0\}$ is a parameter. To
construct solutions of the initial value problem, one may invoke the
so-called WKB ansatz~\cite{dingle1973asymptotic},
\begin{equation} \label{equ_IVPFIO_WKB}
   u_\xi(x,t) = a_+(x,t,\xi) \exp[i \alpha_+(x,t,\xi)]
                  + a_-(x,t,\xi) \exp[i \alpha_-(x,t,\xi)]
\end{equation}
Invoking the initial conditions yields
\begin{equation} \label{equ_IVPFIO_inipha}
   \alpha_+(x,0,\xi) = \alpha_-(x,0,\xi) = \langle \xi,x \rangle
\end{equation}
and
\[
  \partial_t \alpha_{\pm}(x,0,\xi) = \mp c(x) |\xi| .
\]
In the case that the wave speed, $c$, does not depend on $x$ we may
easily find $\alpha_\pm$ and $a_\pm$ explicitly, and the WKB ansatz
gives an exact solution: The eikonal equations are
\[
   \partial_t \alpha \pm c |\partial_x \alpha | = 0, \quad
   \alpha_\pm(x,0,\xi) = \langle x, \xi \rangle
\]
and have solutions
\[
   \alpha_\pm(x,t,\xi) = \langle x, \xi \rangle \mp t c |\xi| .
\]
The transport equations are
\[
   \partial_t a_\pm \pm 2 c \frac{\langle \xi, \partial_x a_\pm \rangle}{|\xi|}
   = 0, \quad a_\pm(x,0,\xi) = \pm \frac{i}{2 c |\xi|} ,
\]
the solutions of which are simply constant
\[
   a_\pm(x,t,\xi) = \pm \frac{i}{2 c |\xi|} .
\]
Thus, in the constant wave speed case the WKB ansatz yields
\[
   u_\xi(x,t) = \frac{i}{2 c |\xi|}
     \left( e^{i (\langle x, \xi \rangle - t c |\xi|)}
       - e^{i (\langle x, \xi \rangle + t c |\xi|)} \right)
   = \frac{e^{i \langle x,\xi \rangle}}{c |\xi|} \sin (t c |\xi|)
\]
It is not difficult to check that when $\xi \neq 0$ this is an exact
solution of the scalar wave equation with the plane wave initial
data.

If we integrate $u_\xi(x,t)$ with respect to $\xi$, then we obtain
oscillatory integrals in $x$ depending on the parameter $t$. The
initial conditions imply that
\[
   u(x,0) = (2 \pi)^{-2} \int u_\xi(x,0) \, \mathrm{d}\xi = 0
\]
and
\[
   \partial_t u (x,0) =
   \partial_t |_{t = 0} (2 \pi)^{-2} \int u_\xi(x,t) \,
            \mathrm{d}\xi = \delta(x) .
\]
Then
\begin{multline} \label{eq:smooth_Greens}
   u(y,t) = (2 \pi)^{-2} \int a_+(y,t,\xi) 
      \exp[i (\alpha_+(y,t,\xi)] \frac{i}{2 c |\xi|} \mathrm{d}\xi
\\
      - (2 \pi)^{-2} \int a_-(y,t,\xi) 
      \exp[i (\alpha_-(y,t,\xi)] \frac{i}{2 c |\xi|} \mathrm{d}\xi ,
\end{multline}
yielding, at fixed time $t$,  the amplitudes and phase functions of two FIOs representing the parametrices of two half wave equations. The canonical transformations, with $t$ fixed, follow from
\begin{eqnarray}
   \frac{\partial \alpha_\pm}{\partial \xi}
                        &=& y \mp t c \frac{\xi}{|\xi|} ,
\\
   \frac{\partial \alpha_\pm}{\partial y} &=& y ,
\end{eqnarray}
yielding $$\biggl(\underbrace{y \mp t c \frac{\xi}{|\xi|}}_{= x},
\xi\biggr) \to (y, \overbrace{\xi}^{= \eta}) ,$$
signifying straight (bi)characteristics.

\section{Training and dataset details} \label{app:training_data_details}

We generate all our data using the MATLAB kWave toolbox\cite{treeby2010k,treeby2018rapid}. We choose our computational domain to be $1024\times 1024$ meters with a grid of size $512\times 512$ keeping the grid spacing at $2$ meters. Our background wavespeeds vary from $1400$ m/s to $4000$ m/s for the reverse time continuation and the inverse source problems and from $2500$ to $4300$ m/s for reflector imaging problem. For simplicity we chose the same background for the reverse time continuation and inverse source problems. For all the problems we choose a propagation time of $T=200$ ms. In order to maintain CFL condition, we need a small time-step. Consequently, our sensor traces in the latter two problems are quite long - with $N > 1500$ data samples in time. To keep the computational burden under control, we subsample all our inputs and outputs to be $128\times 128$ pixels. For the inverse source and reflector imaging problem this represents a subsampling of about $12$x which affect performance. However, we find that we are still able to recover geometry.

We get the spectrally filtered components of our inputs using PyCurvelab~\cite{pycurvelab}. We choose to have $k=4$ scales with $1, 16, 32$ and $1$ wedges respectively per scale. For all our results, we ignore the first and the last scale completely and show results on the middle frequencies. Our method does not change even if we partition all scales. We avoid them here to keep the computational burden under control -- curvelets are very redundant frames, therefore with $N$ boxes in the Fourier space one would convert the input from a single channel input to an $N$ channel input, one per box. A wedge partitioning of the Fourier space as proposed in \cite{candes2006fast} would work in our scheme in terms of learning the geometry, however, theoretically this is not the ideal partitioning for having a sparse separated representation (see Section \ref{sec:fio}). 
 
In all the problems, unless otherwise mentioned the boundary of the domain is modeled via the standard PML (perfectly matched layer) conditions. This means that signals are not reflected back into the domain at the boundaries. 

In order to illuminate a large portion of the domain in the inverse source problem, we reduce our area of interest to $[0,\frac{M}{2}]\times [\frac{M}{4}, \frac{3M}{4}]$ in a domain of size $M\times M$. Note that our sensors are at $\{0\}\times[0,M]$. Similarly for the reflector imaging problem, we reduce our area of interest to $[0,\frac{M}{4}]\times [\frac{3M}{8}, \frac{5M}{8}]$. The reflector imaging problem is more nuanced in that the rays from the reflector are unidirectional as opposed to the inverse source problem where sources propagate in an omnidirectional fashion. Therefore, seismologists would use multiple sources in order to illuminate more orientations in the image space. 

The reverse time continutation problem is tested on dataset of randomly oriented thick lines, random shapes (a mix of circles, rectangles, triangles), randomly rotated MNIST digits, sinusoidal exploding reflectors (inspired from seismics) and Canny edge filtered celebA images~\cite{liu2015faceattributes}.  The inverse source problem is trained on thick lines and tested on the reflectors, shapes and rotated MNIST dataset. For the reflector imaging problem, since this is mainly a seismic imaging technique we keep our datasets restricted to ``layer-like'' inferfaces as seen in the reflectors dataset. To simulate faults, we partition the reflectors dataset into random columns and shuffle them around. We call this shuffled reflectors dataset. Lastly we perform elastic transform on a layered medium and shuffle after partitioning into columns to get more arbitrarily shaped interfaces which we call the random reflectors dataset.

All components of our network are trained by Adam optimizer with a learning rate of $10^{-4}$. The routing network has a learning rate of $10^{-6}$. The U-Net portion of our network has $5$ downsampling blocks and $5$ upsampling blocks in the style of ~\cite{jin2017deep} with $16$ starting channels that double in each donwsampling block. All activations within the U-Net are leaky ReLUs. The downsampling is done via $2$D max pooling. The U-Net takes in all frequencies to allow for channel interaction as per \eqref{equ:Fuy}. Each channel in the output of the UNet is warped as per the grids given by the routing network and then summed to give the final output. In this work, we use $R=1$. We also find that having the multipliers do not signficantly add to the performance of the network. We run our first stage of training for 40 epochs and the second stage for another 80 epochs.

The baseline network is a U-Net with everything the same except it has $6$ downsampling blocks and $32$ starting channels and therefore about $6-10$x more parameters. In our tests, these U-Net networks performed the best. The baseline is trained over 100 epochs of the training data.

All our hyperparameters were tuned only on the reverse time continuation problem based on a validation set of 100 images. The same hyperparameters were used for all three problems. For all the problems, our training set had 3000 images. All results are shown on images never seen by the networks. This is obviously true for out-of-distribution distribution results.

\paragraph{Pretraining of routing network} A randomly initialized routing network outputs degenerate grids where almost all points are close to zero. Post resampling this leads to a almost a constant image as most pixels have been sampled from a small portion of the domain. Therefore, we pretrain our routing networks. 

Obviously we do not know the background wavespeed and hence choose an arbitrary $p=3$ parameter family of radial basis functions. The first two parameters $(z_1, z_2)$ signify the center of the Gaussian and the last parameter, $z_3$ signifies its isotropic standard deviation. We then fix a randomly chosen $\tilde{z}$ when training for downstream imaging applications in Section \ref{sec:exp}. Note that this does not correspond to or is close to the true warped grids. In fact in our experiments, we found that pretraining using constant wave speeds with speeds being within the range of our problem also works well.

\section{Training the canonical relation from ray paths} \label{app:HJ_flow}

In the process of building our architectures, we built a debugging tool which can be used to learn the geometry directly from a dataset of end points of ray-paths which is an interesting inverse problem in itself. For the reverse time continuation problem we solve both half-waves (refer Appendix \ref{app:fio_wave_connect}), for the inverse source problem and reflector imaging problems we solve only the half-wave from the sensor line into the interior of the domain. First, we explain the reverse time continuation problem as that is the simplest to understand due to the source and target wave-packets both being snapshots in time. 

Consider any parametric family of backgrounds, $c_\theta(\cdot): \R^2\mapsto \R, \theta\ \in\ \R^p$, from which we sample $\{z_i\}_{i=1}^{N_c}$ points according to some prior $p_\theta$ (we chose uniform). For each of the sampled backgrounds, we sample $M$ phase-space points $\{(x_l, \xi_l)\}_{l=1}^{M}$ and simulate how wave-packets at this location in phase space would travel under background $c_{z_i}$ by solving the Hamilton flow using the $4^{\mathrm{th}}$ order Runge-Kutta scheme for integration. We note the final locations of these wave packets at time $T$ in phase space as $\{(y_l, \eta_l)\}_{l=1}^{M}$ and thus generate $MN_c$ training pairs $\{(y_i,\xi_i), (x_i, \eta_i)\}_{i=1}^{MN_c}$. Note that we have the \emph{final} location and \emph{initial} orientation as input and \emph{initial} location and \emph{final} orientation as output in accordance with the box algorithm laid out in~\cite{andersson2012multiscale}. We proceed by training a 4 layer fully-connected network $N_T(z, y, \xi)$ that takes in $(y,\xi)$ along with the parameter vector, $z$ and gives the estimate for $(\tilde{x}, \tilde{\eta})$. Note that due to homogeneity of the phase function, $S$ the network only cares about the direction $\hat{\xi}$ and not the magnitude. Our network layer sizes are $(4+p, 32, 64, 128, 4)$ with leaky ReLU activations except for the last layer which has identity activation. We do not employ any normalizing or dropout strategies for training. We use the trained fully connected networks to generate training data for our routing network by calculating how an entire grid, $G$ of wave-packets oriented in direction $\nu$ would warp. This is because routing network outputs entire warped grids in one go using convolutional layers and therefore cannot be trained on ray paths directly. 

\begin{figure}
    \centering
    \includegraphics[width=0.9\textwidth]{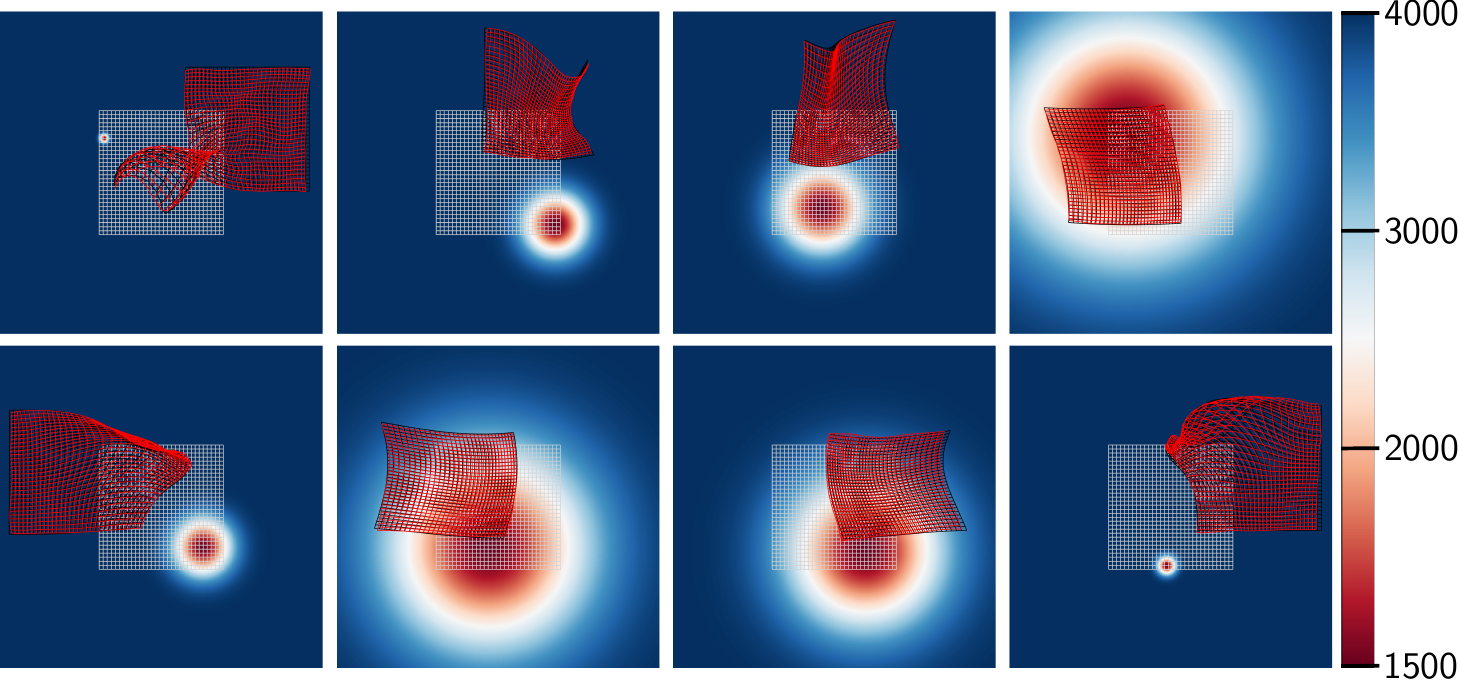}
    \caption{Examples of the learnt coordinate transforms trained directly from ray paths over a 3-parameter family (2 for center location, one for standard deviation) of simple radial basis functions. The red is our prediction while the black grid shows the actual deformed grid calculated using Hamilton flows. Note from the top left and bottom right figures that we can also predict caustics.}
    \label{fig:def_grids}
\end{figure}

We find that such a simple characterization of the canonical relation also allows for caustics to develop as shown in Figure \ref{fig:def_grids} which is usually avoided in the literature on FIOs~\cite{candes2007fast}. 

Note that the training is slightly different for the inverse source and reflector imaging problems. For these problems we need pairs $\{(y_i,(\xi_{1i}, \tau_i)), ((x_{1i}, t_i), \eta_i)\}_{i=1}^{MN_c}$ as we pair each wave-packet seen in the sensor trace $(x_1,t)$ domain to the initial source location in $(y_1, y_2)$ domain. Here $\tau = c_z(x)|\xi| \ \forall\ (x, \xi)$ since rays follow a Hamiltonian system. In reflector imaging, the $(y,\eta)$ point corresponds to the point on the ray right after reflection. 

\end{appendices}

\end{document}